\RequirePackage{fix-cm}
\documentclass[twocolumn,epjc3]{svjour3}  
\smartqed  
\RequirePackage{graphicx}
\usepackage[dvipsnames]{xcolor}
\usepackage{tikz}
    \usetikzlibrary{patterns}

\usepackage{amsmath}
\usepackage{amssymb}
\usepackage{nccmath}
\usepackage{graphicx}
\usepackage{comment}

\usepackage{ragged2e}
\usepackage{subcaption}
\usepackage{enumitem}
\usepackage{booktabs}
\usepackage{float}
\usepackage{hyperref}
\usepackage{multirow}
\usepackage{mwe}
\journalname{Eur. Phys. J. C}
\begin{document}

\title{Optimization of the X-Arapuca Photon Collection Efficiency for the DUNE Horizontal Drift Far Detector}

\author{
    E.~Bertolini\thanksref{addr1,addr2}
    \and
    C.~Brizzolari\thanksref{addr1,addr2}
    \and
    F.~Bruni\thanksref{addr3}
    \and
    P.~Carniti\thanksref{addr1,addr2}
    \and
    C.M.~Cattadori\thanksref{addr1,addr2}
    \and
    S.~Copello\thanksref{addr4}
    \and
    E.~Cristaldo\thanksref{addr1,addr2}
    \and
    M.~Delgado\thanksref{addr1,addr2}
    \and
    F.~Galizzi\thanksref{addr1,addr2}
    \and
    C.~Gotti\thanksref{addr1,addr2}
    \and
    D.~Guffanti\thanksref{addr1,addr2}
    \and
    A.A.~Machado\thanksref{addr5}
    \and
    L.~Malinverni\thanksref{addr4}
    \and
    L.~Meazza\thanksref{e,addr1,addr2}
    \and
    F.~Meinardi\thanksref{addr3}
    \and
    G.~Pessina\thanksref{addr1,addr2}
    \and
    G.~Raselli\thanksref{addr4}
    \and
    M.~Rossella\thanksref{addr4}
    \and 
    E.~Segreto\thanksref{addr5}
    \and
    H.~Souza\thanksref{addr1,addr2}
    \and
    F.~Terranova\thanksref{addr1,addr2}
    \and
    D.~Warner\thanksref{addr6}
}
\thankstext{e}{e-mail: luca.meazza@mib.infn.it}

\institute{
    \textbf{Dipartimento di Fisica “Giuseppe Occhialini”, Università degli Studi di Milano-Bicocca}, Piazza della Scienza 3, 20126 Milano~(Italy) \label{addr1}
    \and
    \textbf{INFN Sezione di Milano-Bicocca}, Piazza della Scienza 3, 20126 Milano~(Italy) \label{addr2}
    \and
    \textbf{Dipartimento di Scienze dei Materiali, Università degli Studi di Milano-Bicocca}, Via Roberto Cozzi 55, 20125 Milano~(Italy)\label{addr3}
    \and
    \textbf{Dipartimento di Fisica “Alessandro Volta”, Università di Pavia}, Via Bassi 6, 27100 Pavia~(Italy) \label{addr4} 
    \and
    \textbf{Instituto de Física “Gleb Wataghin”, UNICAMP}, Campinas-SP, 13083-859~(Brazil) \label{addr5}
    \and
    \textbf{CSU}, Limelight Ave, Castle Rock 80109~(United States of America) \label{addr6} 
}

\date{Received: date / Accepted: date}

\maketitle

\begin{abstract}
The Deep Underground Neutrino Experiment (DUNE) Far Detector (FD) Photon Detection System (PDS) employs the X-Arapuca concept, a photon trapping system relying on reflective surfaces and dichroic filters. In this paper are reported measurements, performed at the University of Milano-Bicocca, aimed at increasing the FD Horizontal Drift (HD) PDS module efficiency. The baseline implementation of the X-Arapuca concept for the FD-HD PDS module is close to the DUNE requirements as demonstrated in the collaborations laboratory testing. However, an increased performance would provide a safety margin for a detector planned to be operated for 30 years, without possibility of performing maintenance. A higher detector performance would also benefit the DUNE low energy physics program.\\
The already proven Milano-Bicocca setup has been utilized to test different PDS module configurations comparing them to the original baseline. Exploiting prior knowledge of the X-Arapuca components and Geant4 based optical simulations it has been possible to achieve up to an $\sim$84\% performance increase over the baseline design. In the following it is presented the testing procedure, the performed measurements and a brief discussion on the obtained results.

\end{abstract}

\setcounter{tocdepth}{2}
\tableofcontents

\section{Introduction}
The Deep Underground Neutrino Experiment (DUNE) is a long baseline neutrino experiment with the goal of performing a precision measurement of the oscillation parameters in order to determine the neutrino mass hierarchy and test the CP symmetry violation in the weak sector \cite{TDR1_1807.10334,TDR2_2002.03005}. The experiment will be composed of a powerful neutrino beam and a Near Detector, both located in Illinois, and a modular Far Detector (FD), located 1.5~km underground in South Dakota, 1300~km from the neutrino source. The first two modules of the FD will be 17~kton Liquid Argon Time Projection Chambers (LArTPCs), reading both charge and light generated by charged particles ionizing the argon atoms, in order to reconstruct 3D particle tracks with millimeter precision. The scintillation light is read by the Photon Detection System (PDS) whose modules utilize the X-Arapuca concept to trap and detect photons.\\
This paper presents the work done to improve the photon-collecting efficiency of the DUNE FD-HD X-Arapuca baseline module. 
The Photon Detection Efficiency (PDE) of the baseline module, installed and under test in ProtoDUNE~\cite{Álvarez-Garrote2024}, satisfies DUNE's physics requirements; however, a further improvement of its PDE would provide a safety margin over the minimum required specifications (i.e. in case of degradation over the detector lifespan) and an enhanced sensitivity of the PDS for low-energy physics.\\
The PDE enhancement work focused on identifying the factors that limited the baseline module performance and on finding possible solutions to improve it while maintaining the design as close as possible to the baseline in order to minimize changes to the mechanical frame.
The mechanical constraints were driven by the need for an easy-to-implement solution for the measurement in the Milano-Bicocca setup; the mechanical configuration described in this paper is therefore to be considered as a proof of concept while the implementation for the DUNE module might vary depending on the engineering requirements. The physical variables relevant to the photon-collecting efficiency will be discussed in the following sections.\\
\section{The X-Arapuca}
\begin{figure}
    \centering
    \includegraphics[width=0.5\linewidth, angle=-90 ]{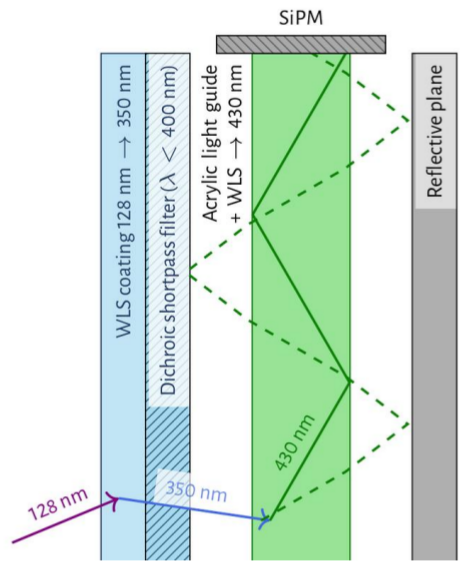}
    \caption{The X-Arapuca operating principle~\cite{Guffanti}.}
    \label{fig:schemaXA}
\end{figure}

The Photon Detection System of the DUNE Far Detector LArTPC is based on the X-Arapuca (XA) technology \cite{Machado2016,Machado2018}. The XA allows to detect the Vacuum Ulta Violet (VUV) LAr scintillation light ($\lambda \sim 128~\rm{nm}$) with commercially available and cryo-reliable photosensors; this type of photosensor is currently available with a detection efficiency centered in the visible light range.
The XA is a reflective box that features an entrance window, two wavelength shifting stages, a dichroic filter that allows light to enter but not to escape the device and a light guide that collects and coveys the light onto Silicon Photomultipliers (Fig.~\ref{fig:schemaXA}).
The first downshift, from 128~nm to 350~nm, is performed by a wavelength shifting (WLS) coating of paraterphenyl (pTP) deposited on a dichroic filter that covers the entrance window of the device. The photons are downshifted again by the wavelength shifter light guide (WLS-LG), from 350~nm to 430~nm, and collected onto Silicon Photomultipliers (SiPMs). The photons that escape the light guide are kept inside the XA by the reflective surfaces (lined with Vikuiti \cite{3M}) and the dichoic filter on the entrance window, which has a cut-off at 400~nm until they are detected from the SiPMs or absorbed by the materials.\\
The FD-HD PDS is composed of 1500 modules, each of which contains four equal channels, therefore, laboratory measurements are performed on a single channel. Each channel has dimensions of $\sim$50~cm$\times$10~cm$\times$1~cm, with 48 SiPMs distributed evenly along the long edges, 24 per side. The main features of the FD1-HD and FD2-VD X-Arapuca design are reported in Table~\ref{tab:XA_par} 

\begin{center}
    \begin{table}
    \small
    \centering
    \begingroup
    \renewcommand{\arraystretch}{0.9}
        \begin{tabular}{ c c c }
        \toprule
         & \textbf{HD} & \textbf{VD} \\
        \midrule
        \textbf{WLS-LG size[$mm^2$]} & {480 × 93} & {607 × 607} \\
        \textbf{WLS-LG surface [$m^2$]} & $0.04 $ &  $ 0.36$ \\
        \textbf{N. SiPMs} & 48 & 160 \\
        \textbf{SiPMS coverage} & $3.8\%$ & $1.7 \%$ \\
        \textbf{PDE} & $2.5\% - 4.6 \% $ & $3.7\% - 4.5\%$ \\
        \textbf{FOM} & $ 0.65 -1.2 $ & $ 2.4 - 2.9 $ \\
        \bottomrule
        \end{tabular}
    \endgroup
    \caption[VD and HD PDS modules] {HD and VD PDS module specifications comparison. Here: SiPM coverage = (N. SiPM x SiPM size (6 × 6 $mm^2$)) / WLS-LG size and FOM = PDE / SiPM coverage.}
    \label{tab:XA_par}
    \end{table}
\end{center}
%
\subsection{The entrance window} 
\begin{figure}
    \centering
    \includegraphics[width=1.0\linewidth]{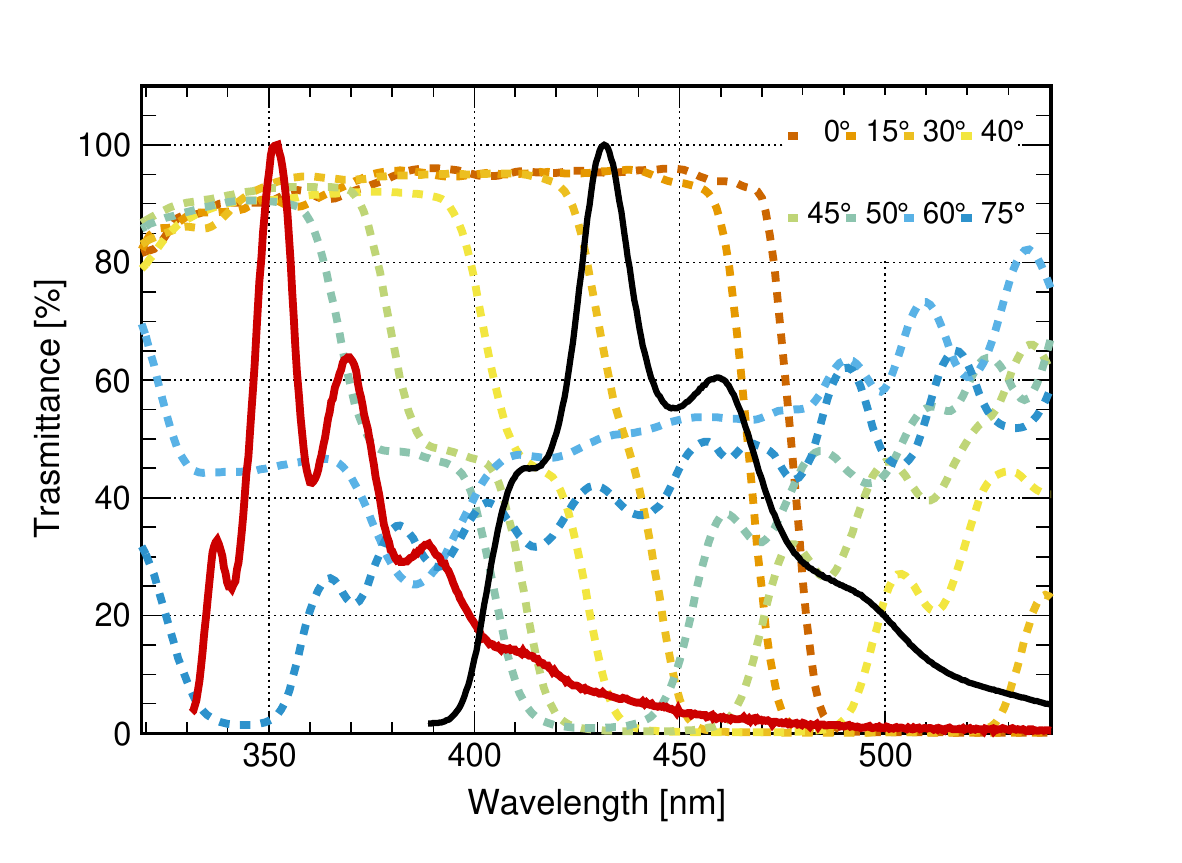}
\caption{ZAOT DF Transmission curves measured in water ($n_{H2O}=1.33$)\cite{Cattadori_2024}. The two PL spectra of the pTP (red) and of the secondary WLS chromophore (gray) embedded in the PMMA are also reported}.
    \label{fig:ZAOT_TC}
\end{figure}

The X-Arapuca entrance window is a high trasmittance BF33 glass slab from Schott, pTP is vacuum evaporated on the external side with a target thickness of about 400~$\mu\rm{g/cm}^2$; depending on the deposition run and position in the evaporator, the achieved pTP thickness ranges from 280 to 600~$\mu\rm{g/cm}^2$.
The choice of the pTP thickness has been validated by a set of measurements, taken with the apparatus described in Section \ref{sec:Pavia}, showing that for thicknesses greater than 200~$\mu\rm{g/cm}^2$ a plateau on the conversion efficiency is reached.
On the internal side, a dichroic multilayer thin film interferometer (DF) is coated. It is designed to transmit the 350 nm light emitted by pTP, and to reflect the 440 nm light emitted by the WLS-LG that escapes its surfaces when impinging above the critical angle $\theta_c = 56^o = \arcsin{(n_1/n_2)}$, where $n_1$=1.24 and $n_2$=1.49 are the LAr and PMMA refraction indices respectively. Dichroic filters are designed for a precise cutoff wavelength $\lambda_c$ and for a narrow range of angles of incidence (AOI). Figure \ref{fig:ZAOT_TC} shows the transmission curves ($T_C$) of the dichroic multilayer thin film coating custom designed, under our guidance, for $\lambda_c =400~\rm{nm}$ and AOI $45^o$ in LAr (the green curve), coated on Schott BF33 glass substrates, by the ZAOT Company (Italy) \cite{ZAOTco} for the FD2 X-Arapuca~\cite{Cattadori_2024}; the measurements are performed with the Essentoptics Photon RT spectrophotometer in a vial filled  with ultrapure water ($n_{H2O}=1.33$), the best proxy of LAr at room temperature. These DF were deployed in ProtoDUNE Vertical Drift (NP02) at the CERN Neutrino Platform. Figure \ref{fig:ZAOT_TC} also shows the  Photo Luminescence (PL) curve of the primary pTP and secondary WLS chromophore embedded in the PMMA, measured at 77 K with a TM-C10083CA Hamamatsu Mini-Spectrometer; these measurements have been performed at the Milano-Bicocca department of Material Science.\\
The DFs exhibit good transparency ($>$ 85\%) for the primary radiation at 330-380~nm for AOI from $0^o$ to 50$^o$: for larger AOI  their transmittance is largely compromised. As pTP radiation is emitted following a Lambertian distribution, a large fraction is filtered out, it has no chance of entering the box, to be downconverted and  transported to the photosensors.
In addition, Figure \ref{fig:ZAOT_TC} shows that the DF reflection dip, which is designed to reflect back the 430~nm radiation escaping the WLS-LG for $\theta$ \textgreater 56\textdegree, adequately matches the PL spectra of the secondary WLS for 30\textdegree \textless  AOI  \textless 50\textdegree. Outside this range, only a (minor) fraction is efficiently reflected back into the box by the DF, the rest being transmitted and leaving the X-Arapuca with high probability. The overall impact depends on the X-Arapuca device's geometry and size, as discussed in Section \ref{sec:entrancewindow}. In this work we tested the DF performance in LAr once embedded in a FD1 X-Arapuca device as described in Sections~\ref{setup}~and~\ref{data}.

\subsection{Assessment of the entrance window  transparency}\label{sec:Pavia}

\begin{figure}
    \centering
    \includegraphics[width=1.0\linewidth]{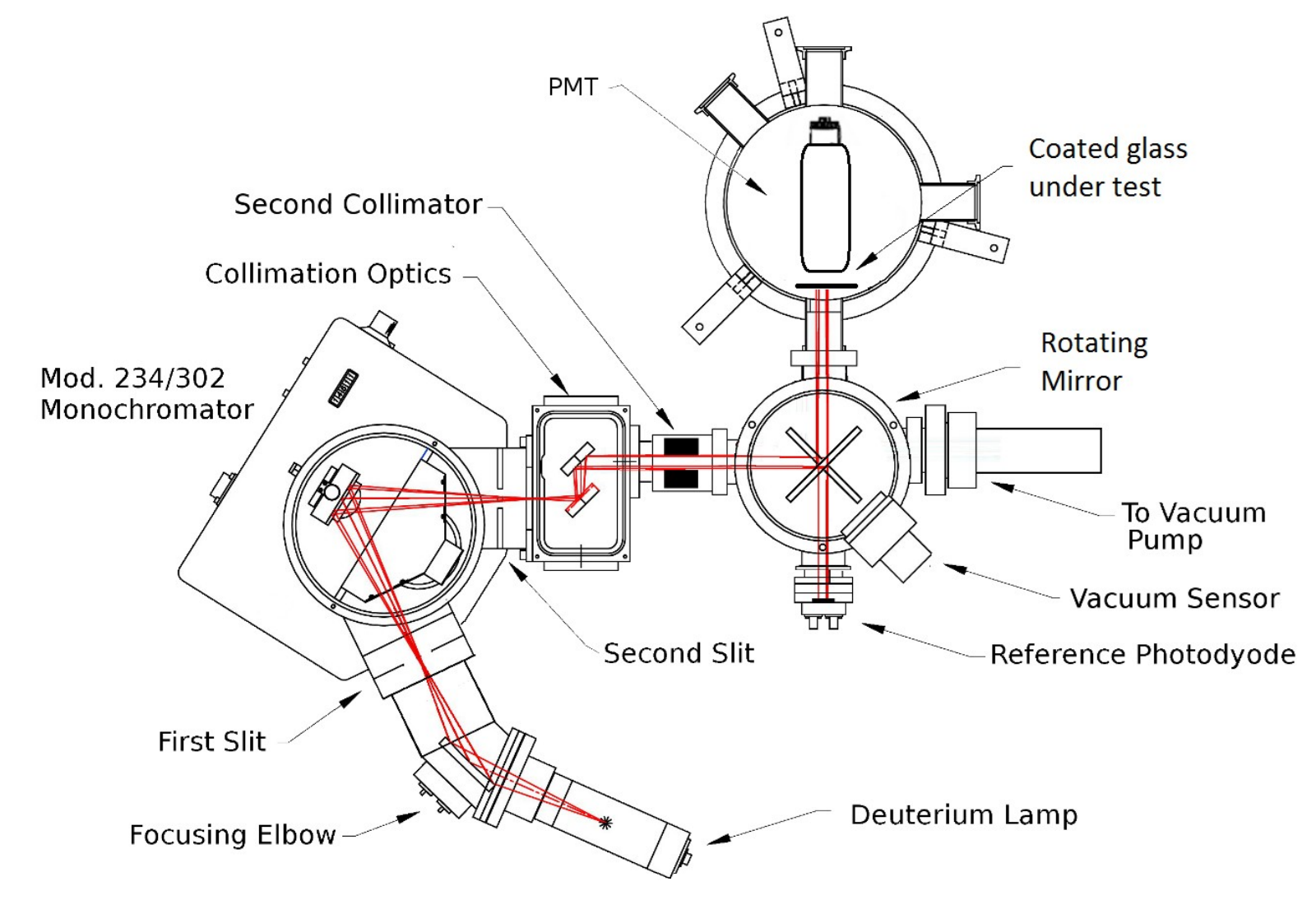}
\caption{A schematic representation of the vacuum setup used to evaluate, at room temperature, the conversion and transmission efficiency of different window samples.}.
    \label{fig:pavia}
\end{figure}

Given the isotropic emission of the pTP, and the DF transparency dependence on the incoming light angle of incidence and wavelength, the overall DF transparency to incoming pTP light is given by the equation:%
\begin{equation}
    \gamma_{pTP}^{trans}=\int_{\theta}\int_{\lambda}T_{DF}(\theta,\lambda)\times\gamma_{pTP}^{inc}(\theta,\lambda)\ d\theta\ d\lambda
\label{eq:entrance_window_trasparency}
\end{equation}%
The amount of photons being transmitted ($\gamma_{pTP}^{trans}$) is given by the integral of the distribution of incident photons coming from pTP ($\gamma_{pTP}^{inc}(\theta,\lambda)$) weighted by the DF transmittance ($T_{DF}(\theta,\lambda)$) over all the incidence angles $[0,\pi/2)$ and wavelengh range of the pTP emission spectrum. A custom geant4 Montecarlo simulation, taking as input the DF transmittance curves and the optical properties of the entrance windows materials, has shown that the DF coating can significantly lower the transparency for incoming VUV photons.\\
To directly quantify the overall efficiency of both the pTP wavelength conversion and DF transmission, a set of measurements was performed using windows $97\times 97~\rm{mm}^2$ in size. Here, the efficiency is defined as the number of photons that are down-converted and passing through the inner face of the entrance window over the number of primary VUV photons hitting the window. These measurements were performed in a vacuum setup, shown in Figure~\ref{fig:pavia}, described as follows.

A Hamamatsu L2D2 Deuterium Lamp provides the primary VUV light which, once focused, passes through a McPherson monochromator with a 120 lines/mm grating to select the 128 nm component. A collimator reduces the beam diameter to approximately 8 mm. An intermediate chamber contains a rotatable aluminum mirror that redirects the beam either toward the sample chamber or the intensity monitor. The latter is a calibrated photodiode that measures the flux of primary photons and compensates for lamp instabilities and aging. This is achieved by measuring the primary light both before and after interaction with the sample. The sample chamber houses the device under test (one X-Arapuca entrance window), placed in front of a photomultiplier tube (PMT) model R6091 MOD by Hamamatsu. The PMT, operated at low gain, is used to measure the intensity of the down-shifted light.
The entire system operates under vacuum (around $10^{-4}$ mbar) to reduce VUV absorption to a negligible level.
The absolute value of the conversion efficiency is obtained by comparing the PMT current measured with and without the device under test, and correcting by the ratio of the PMT quantum efficiencies (as declared by Hamamatsu) at the primary 128 nm and the down-shifted light.\\
Two types of measurements were performed: sampling different portions of the same window (to assess uniformity), and sampling different windows (to assess homogeneity across different samples).
For the uniformity measurement, five different spots on the same sample were investigated. The resulting efficiencies ranged from 15\% to 22\%, with a mean value of $\epsilon_{\text{singleDF}} = 20\%$ and a 1-$\sigma$ spread among the measurements of $\sigma_{\text{singleDF}} = 4\%$.
The homogeneity measurement was performed by sampling the central region of 14 different windows (from the ZAOT batch), resulting in a mean efficiency of $\epsilon_{\text{multipleDF}} = 15\%$ and a standard deviation of $\sigma_{\text{multipleDF}} = 2\%$ (measurements spread).\\
In addition, to check for the impact of the pTP coating aging, a final set of measurements was performed on a re-coated window. One window from the ZAOT batch was cleaned and re-coated with pTP at the INFN Pavia laboratory. Its efficiency was measured in five different spots. In this case, the efficiency was slightly higher: $\epsilon = 22\%$ with a standard deviation of $\sigma = 2\%$.
This last result has been used to obtain a rough estimation of the transmission efficiency of the DF alone. Indeed the same 5-spots measuring procedure has been done with samples coated with pTP and no DF: here a conversion efficiency of $\epsilon_{\text{noDF}} = 30\% \pm 2.4\%$, so that we can conclude that about $(30\%-22\%)/30\% = 27\%$ ($\pm 12\%$ given the uncertainties on the mean values used here) of the shifted light is lost because of the presence of the DF.
\section{The Liquid Argon test bench\label{setup}}
The setup used for the measurements in LAr has been described in \cite{PDE_XA_JINST,Álvarez-Garrote2024}, with the addition of a LED light source to improve the photo-detector calibration.
\begin{figure}
	\centering
	\begin{subfigure}{0.23\textwidth}
      \centering
	   \includegraphics[width=0.7\linewidth]{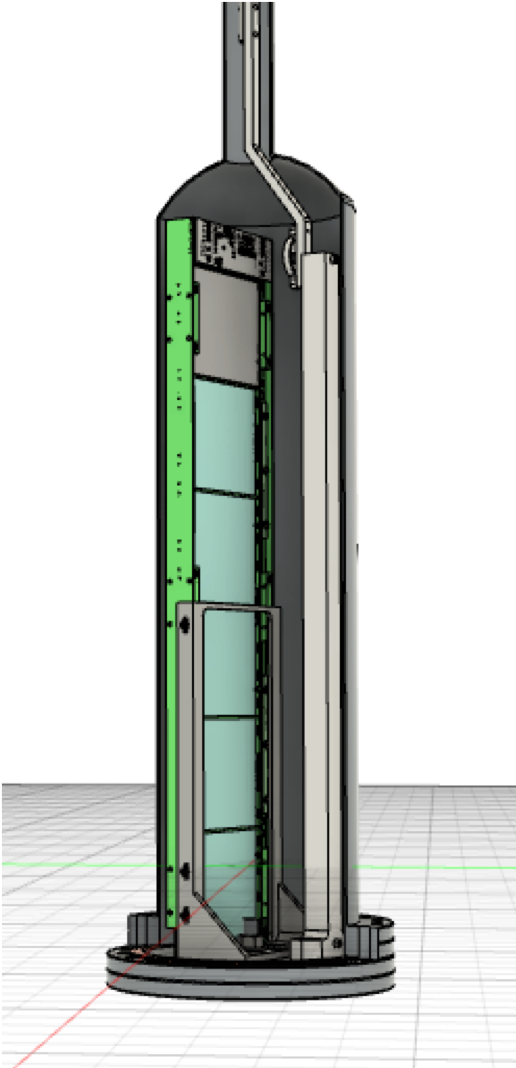}
	\end{subfigure}%
	\begin{subfigure}{0.23\textwidth}
      \centering
	   \includegraphics[width=1\linewidth]{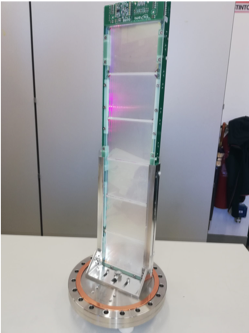}
	\end{subfigure}
	\caption{CAD model of the XA-HD module channel inside the setup vacuum vessel and picture of the channel mounted on the vessel flange.}
	\label{fig:setup_mib}
\end{figure}

    \subsection{Setup}
    The device under test is positioned in a stainless steel $\sim$10~l cylindrical chamber of 150~mm diameter and 550~mm height (Fig.~\ref{fig:setup_mib}). 
    The chamber is vacuum-tight closed and placed in an open 70~l dewar, vacuum pumped down to $O({10^{-4})}$~mbar and then connected to a bottle of 6.0 grade gas Argon (GAr), the dewar is then filled with LAr.
    The GAr is allowed to flow inside the chamber and is liquefied at the expense of the external LAr bath. Once the LAr level inside the chamber fully covers the front end readout circuit, above the XA, the liquefaction process is stopped.
    An exposed $^{241}\rm{Am}$ $\alpha$--source (3.7~kBq) can slide in a vertical rail  by means of a magnetic actuator at the distance of ($55\pm1$~\rm{mm})  from the device entrance window, so that the PDE can be measured at any point along the z-axis.
    Data acquisition is performed using a CAEN DT5725 digitizer configured in self--trigger mode, with rates of approximately 1~kHz and 100~Hz for the $\alpha$--particles and the muons runs, respectively. This digitizer features a 14-bit resolution and a 250~MHz sampling rate (one sample, or \textit{tick}, corresponds to 4~ns).

    \subsection{SiPMs and Electronics}
    The detector has been equipped with the same SiPMs for all the tested configurations and read out with the same electronics as in \cite{Álvarez-Garrote2024}. The SiPMs are Hamamatsu Photonics K.K. (HPK) S13360–9935 with 75 $\mu$m cell pitch and High Quenching Resistance (HQR), the active surface area of each SiPM is 6$\times$6 mm$^2$. This SiPM model is the one selected from the DUNE Collaboration for instrumenting the Far Detector modules.\\
    The 48 SiPMs in the tested detector are connected in parallel, and to the input of a transimpedence amplifier called \textit{cold-amplifier} placed on top of the detector, inside the cylindrical chamber, submerged in liquid argon \cite{Brizzolari_2022}. The amplifier is based on a SiGe bipolar transistor (Infineon BFP640) followed by a fully differential operational amplifier (Texas Instruments THS4531), and is designed for low noise at low power, giving a voltage white noise density of 0.37 nV/$\sqrt{Hz}$ at 2.4 mW per channel. At room temperature, the so–called warm–electronics converts the differential signal to single ended and introduces a second amplification factor, to better match the dynamic range of the analog to digital converters.
\section{Data acquisition, analysis and simulation}
The data acquisition and analysis performed for the measurements reported in this paper are mostly unchanged from the previous campaigns and extensively described in \cite{Álvarez-Garrote2024}, we utilized ROOT~\cite{ROOT} as analysis software. The main differences are a new calibration system utilizing an external LED light source and the convolution approach for the LAr purity analysis (in place of the previous deconvolution approach). The testing procedure is briefly described in the following.
    
    \subsection{Calibration}
\begin{figure}
        \centering
        \begin{subfigure}{0.23\textwidth}
             \includegraphics[width=1\linewidth]{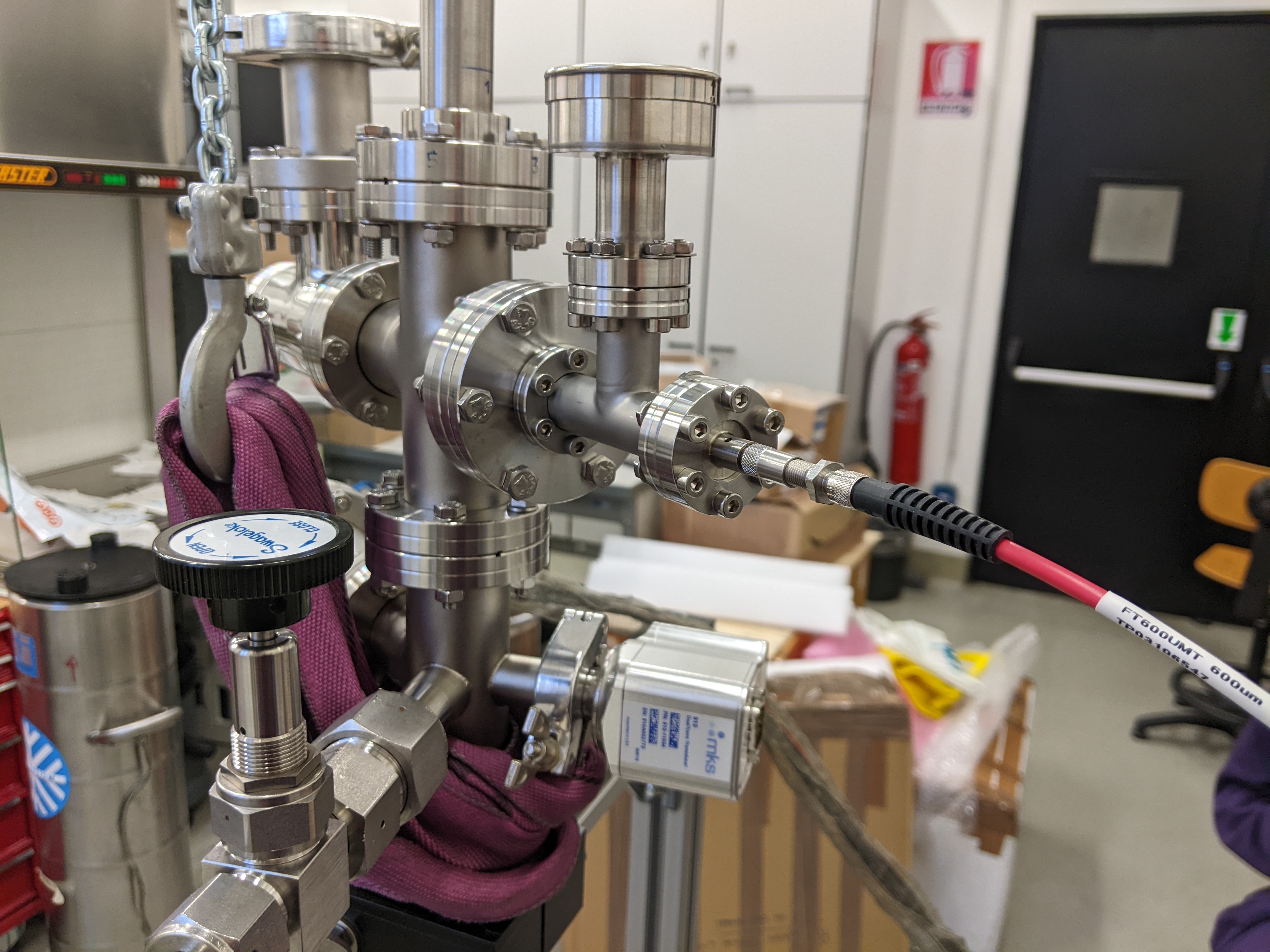}
        \end{subfigure}
        \begin{subfigure}{0.23\textwidth}
            \includegraphics[width=1\linewidth]{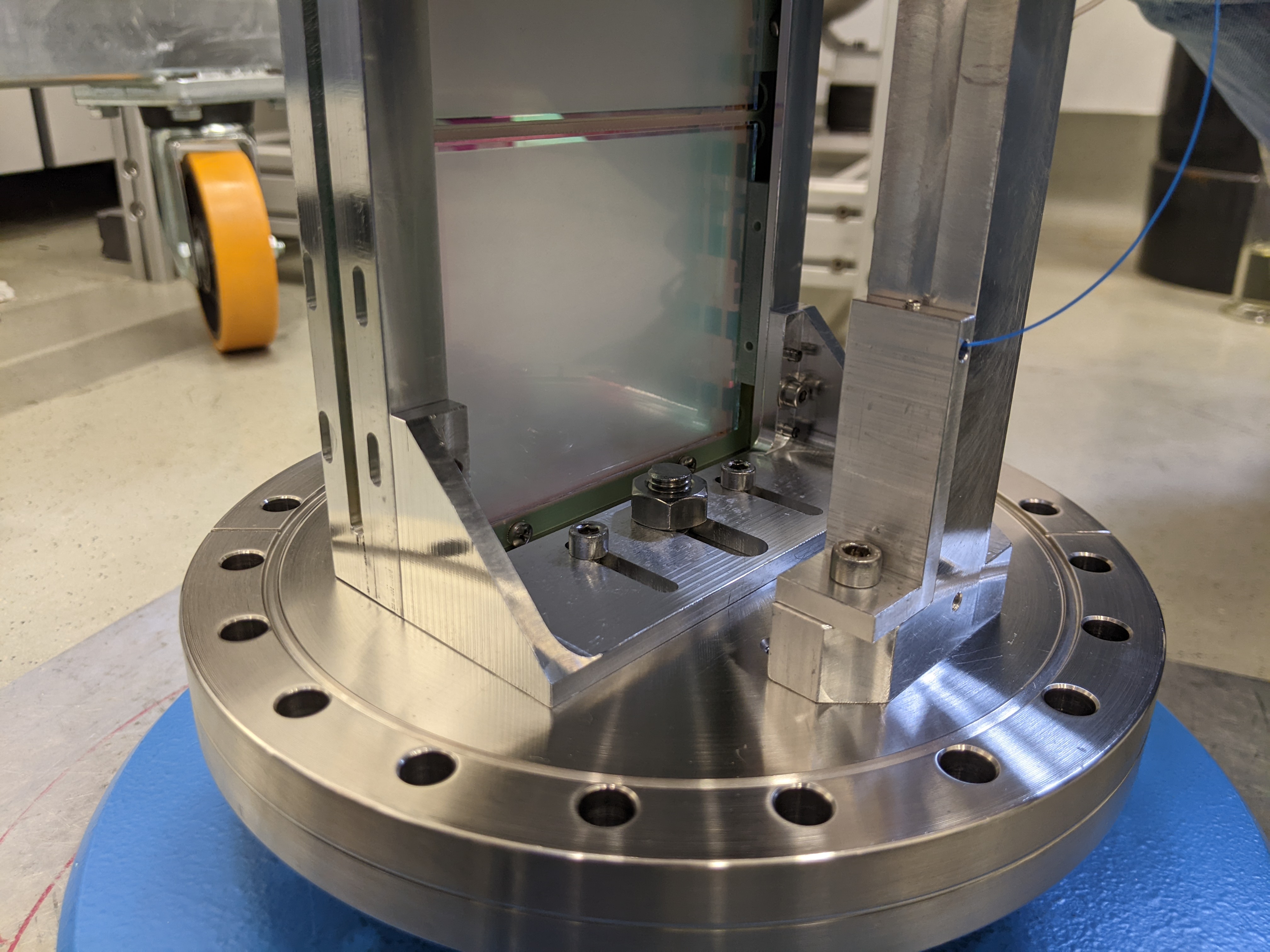}
        \end{subfigure}
        \caption{On the left, optical fiber feed through at the top of the chamber. On the right, optical fiber end fixed to the support next to the alpha source rail.}
        \label{fig:OFft}
    \end{figure}

\begin{figure}
        \centering
        \begin{subfigure}{0.45\textwidth}
            \includegraphics[width=1\linewidth]{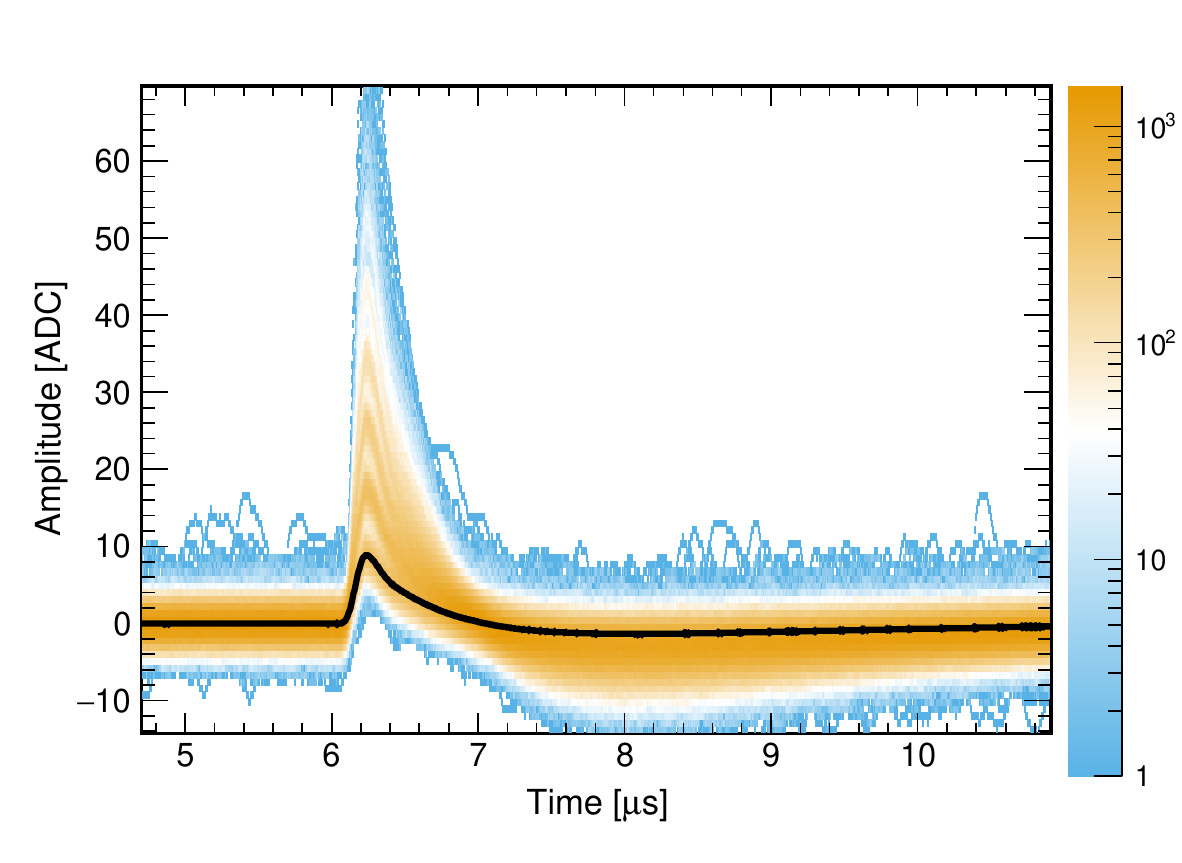}
        \end{subfigure}\\
        \begin{subfigure}{0.45\textwidth}
            \includegraphics[width=1\linewidth]{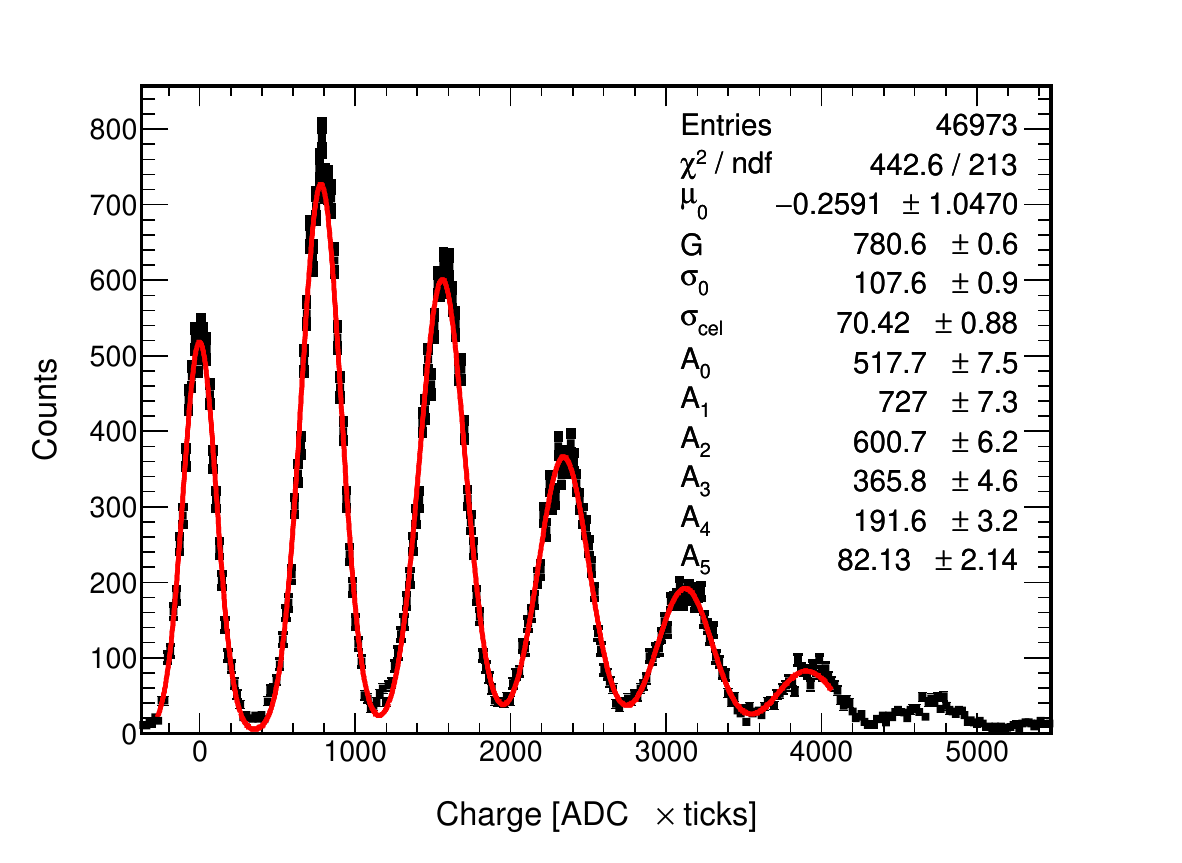}
        \end{subfigure}\\
        \begin{subfigure}{0.45\textwidth}
            \includegraphics[width=1\linewidth]{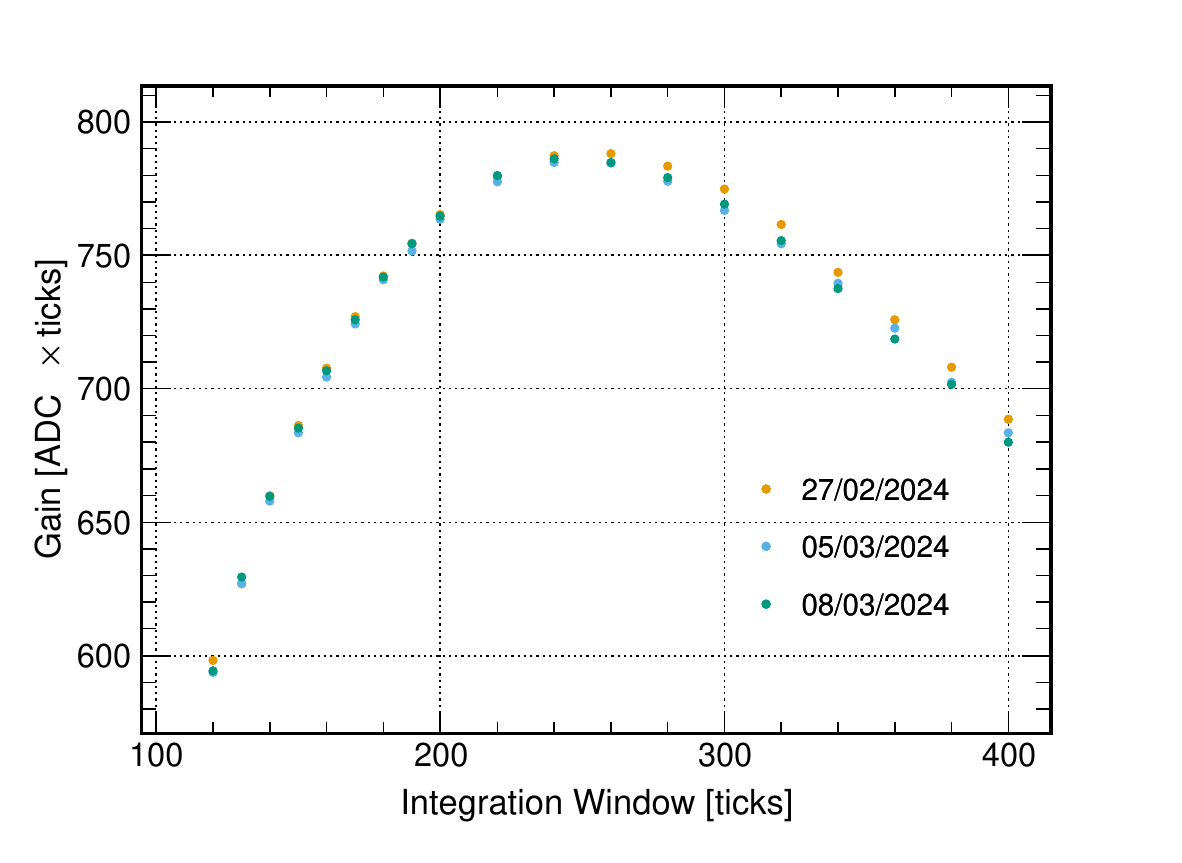}
        \end{subfigure}
        \caption{Top: s.p.e. pulse. Middle: charge spectrum of a calibration run triggered by the external LED. Bottom: gain as a function of the integration window and its stability in time for different runs and filling of the experimental chamber.}
        \label{fig:Gain}
    \end{figure}

    To retrieve the number of detected photo electrons when irradiating the device with the alpha source, the alpha pulses are calibrated by the charge of the average single photoelectron pulse (Fig.~\ref{fig:Gain} top panel). In our previous works \cite{PDE_XA_JINST,Álvarez-Garrote2024}, the single photoelectron events were searched in the events pre-trigger by using a peak-finder algorithm. For this work the vacuum vessel has been equipped with a cryogenic optical fiber, one vacuum tight optical feedthrough and an external 410~nm LED source shown in Figure~\ref{fig:OFft}. The LED driver provides the trigger pulse for the acquisition, thus simplifying the data analysis and reducing the calibration uncertainty.\\
    Calibrations are performed before and after each alpha scanning to check for the stability of both the gain and the Signal-to-Noise ratio (S/N).
    Figure~\ref{fig:Gain} (top), shows the persistence graph of the 3000 events: the average single photoelectron (s.p.e.) pulse is overlapped.
    The integral of the pulses' positive lobe populates the charge histogram of Figure~\ref{fig:Gain} (middle). Thanks to the low noise of both the front-end electronics and of the setup, several peaks show up in the charge spectra each of them corresponding to the detected number of photoelectrons.  The charge spectra is fitted with a series of equally spaced gaussian functions, the distance between them determines the gain, while the ratio of the standard deviation of the s.p.e. peak to its mean value ($\sigma_1/\mu_1$) provides the S/N qualifier.  Figure~\ref{fig:Gain} (bottom) shows that a 250~ticks (1000~ns)  integration window maximizes the pulse positive lobe charge collection, hence we adopt this value; additionally it shows the superior gain stability of our setup across different fillings and data taking that allows to compare results with different configurations of the X-Arapuca and to assess the impact of the individual  components and of their couplings.    

    \subsection{Muon analysis}
\begin{figure}
        \centering
        \begin{subfigure}{0.5\textwidth}
            \includegraphics[width=1\linewidth]{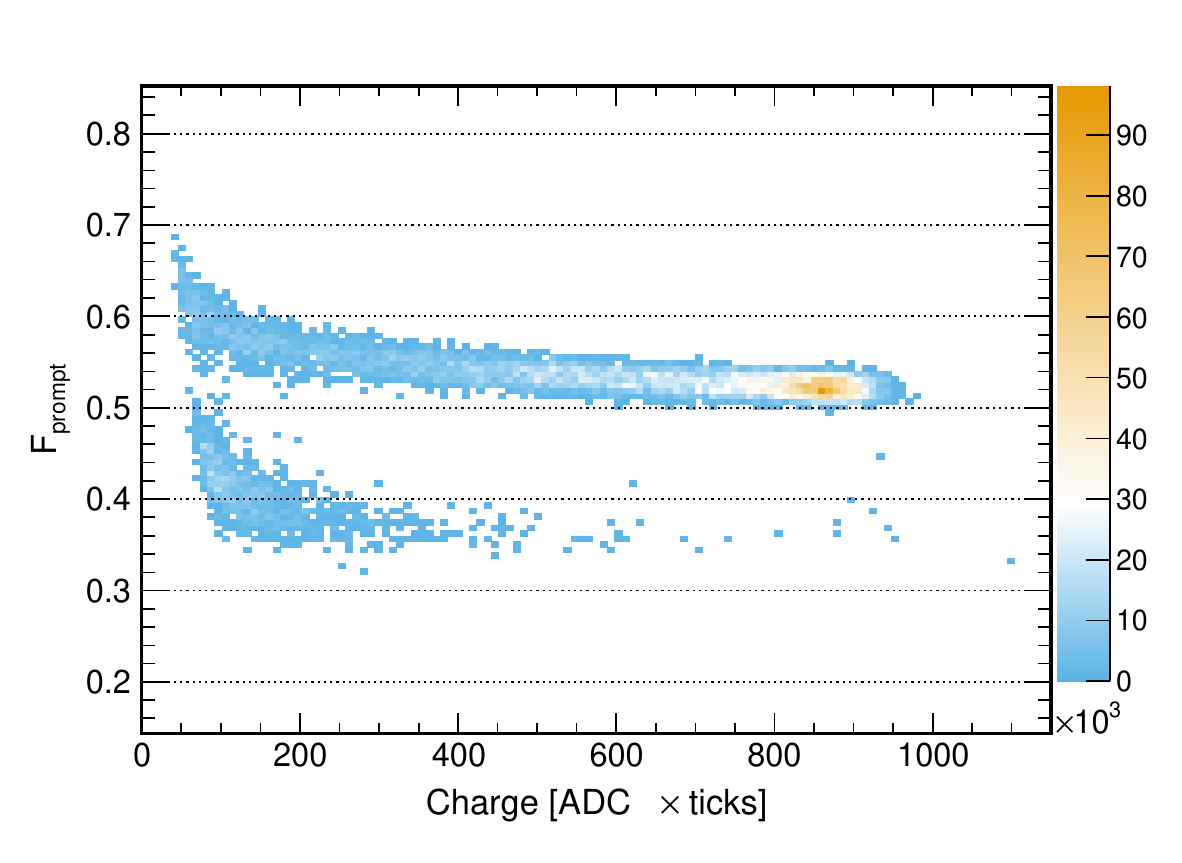}
        \end{subfigure}\\
        \begin{subfigure}{0.5\textwidth}
            \includegraphics[width=1\linewidth]{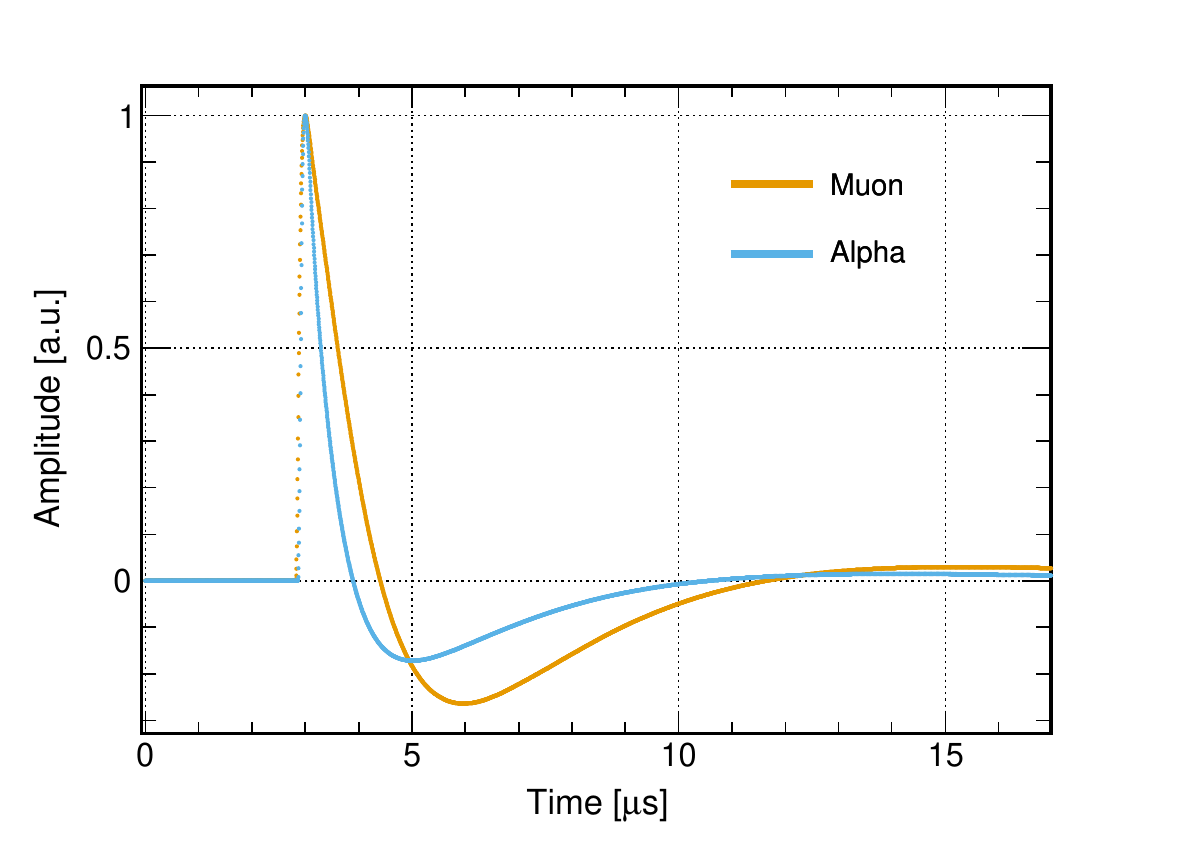}
        \end{subfigure}
        \caption{Top: $F_{prompt}$ vs Charge distribution for an alpha run. The alpha and muon populations are well separated by a 0.5 cut-off in $F_{prompt}$. Bottom: example of alpha and muon average-waveforms (normalized in amplitude).}
        \label{fig:MuonAlphaSelection}
    \end{figure}

\begin{figure}
    \centering
    \includegraphics[width=0.98\linewidth]{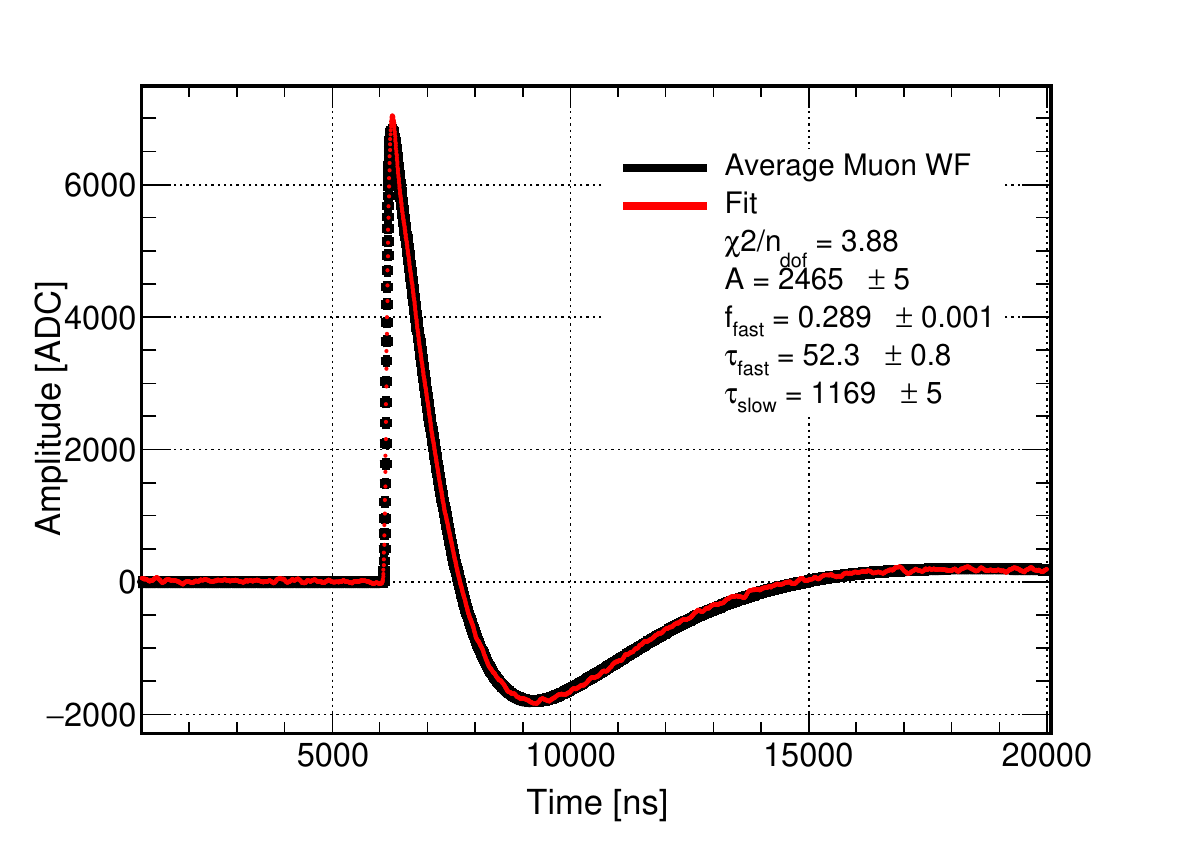}
    \caption{Example of convolution fit \ref{eq_convolution}. The plot refers to the 20 March 2024 run. The legend shows only the fit uncertainties.}
    \label{fig:convolution_fit}
    \end{figure}

    To correct for the liquid argon light-yield loss due to impurities \cite{Acciarri_2009_N2+O2inLAr}, we estimated the liquid argon triplet lifetime for each filling. This was done by convolving the system impulse response (\textit{Template}) with a double-exponential distribution \textit{I(t)} representing the liquid argon scintillation time profile, and fitting the average muon-candidate waveform ($\overline{WF}_{\mu}$). We constructed the template by averaging the calibration waveforms in which one or more photoelectrons where detected. Each waveform also had to satisfy specific selection criteria (e.g., no events coincident with the LED signal). Finally, we normalize the average to the single photoelectron amplitude.
    \begin{equation}
    \label{eq_convolution}
        \overline{WF}_{\mu}(t)=Template(t)\ast I(t)
    \end{equation}
    where
    \begin{equation}
    \label{eq_scintillation}
        I(t)=A\left(\frac{f_{fast}}{\tau_{fast}}e^{\frac{t_0-t}{\tau_{fast}}}+ \frac{1-f_{fast}}{\tau_{slow}}e^{\frac{t_0-t}{\tau_{slow}}}\right)
    \end{equation}
    We acquired dedicated runs with the alpha source was lifted above the liquid argon level to minimize the number of triggers due to alpha events. The offline analysis starts with an event selection to discard the waveforms containing more than one pulse within the acquisition window and all the alpha candidates based on the prompt fraction (\textit{$F_{prompt}$}\footnote{$F_{prompt}$ is defined as the ratio of the signal integrals evaluated on a 600~ns (prompt) and a 1000~ns window.}). See Figure \ref{fig:MuonAlphaSelection}.
    With these criteria we aim to obtain the best waveform representing a signal due to scintillation light induced by a crossing muon. We abandoned the deconvolution technique (used in \cite{Álvarez-Garrote2024}) in favour of the convolution to get rid of a free parameter ($\sigma$ in the previous article) and the additional low-pass filter that the deconvolution required. Then we perform the convolution fit (Eq.~\ref{eq_convolution}) with \textit{A}, \textit{$f_{fast}$}, \textit{$\tau_{fast}$}, and \textit{$\tau_{slow}$} as free parameters for 30 $t_0$ values. The scan over $t_0$ is done to take into account the time discretization due to the sample rate. Among the 30 corresponding fits, we take the one with the minimum $\chi^2$ (Fig. \ref{fig:convolution_fit}).
     The liquid argon purity was high ($\tau_{slow} >$1000~ns), with values consistent among all fillings within 50~ns. Table \ref{tab:Ttriplet} reports the $f_{fast}$ and $\tau_{slow}$ results.
    \begin{table}
        \centering
        \begin{tabular}{ccc}
        \toprule
        Measurement & Fast fraction [\%]   & $\tau_{slow}$ [ns]\\
        \midrule
        27/02/2024 & 29 & 1249 \\
        28/02/2024 & 29 & 1108 \\
        \midrule
        04/03/2024 & 31 & 1163 \\
        05/03/2024 & 30 & 1196 \\
        \midrule
        07/03/2024 & 29 & 1263 \\
        08/03/2024 & 29 & 1295 \\
        \midrule
        12/03/2024 & 29 & 1278 \\
        13/03/2024 & 28 & 1314 \\
        14/03/2024 & 28 & 1360 \\
        15/03/2024 & 29 & 1313 \\
        \midrule
        19/03/2024 & 31 & 1168 \\
        20/03/2024 & 30 & 1198 \\
        \midrule
        09/04/2024 & 31 & 1140 \\
        10/04/2024 & 31 & 1137 \\
        \bottomrule
        \end{tabular}
        \caption{Liquid argon scintillation triplet time constant computed for different measurements. We note that the results at each filling (consecutive days) are consistent considering a 1\% and a 50~ns uncertainties on the fast fraction and $\tau_{slow}$, respectively.}
        \label{tab:Ttriplet}
    \end{table}

    \subsection{Alpha analysis}
\begin{figure}
        \centering
        \includegraphics[width=\linewidth]{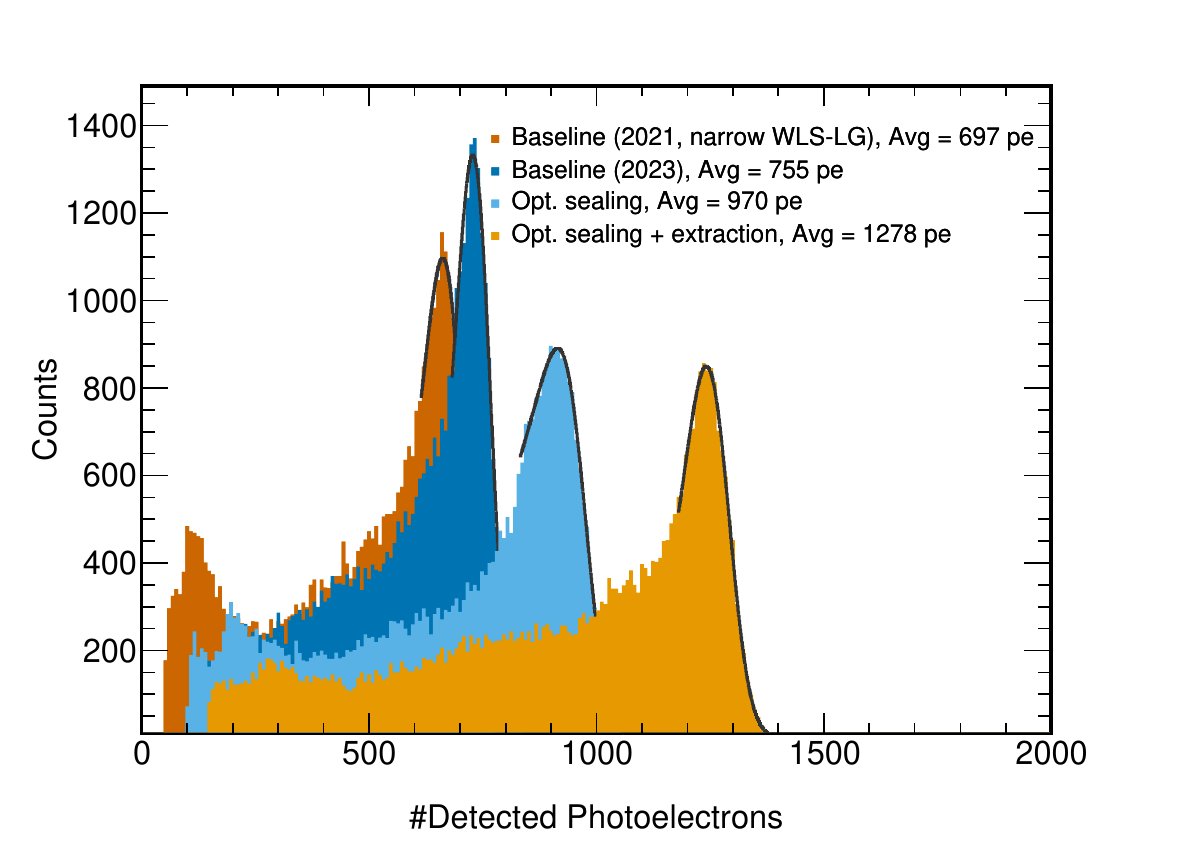}
        \caption{Alpha charge spectra for different module configurations with the same source position (same geometrical acceptance).}
        \label{fig:alpha_spectra}
    \end{figure}

     Thanks to the spectroscopic performance and the high LAr purity achievable with our setup, we use $^{241}$Am $\alpha$ particles to determine the photon detection efficiency of the device under test. A 3.7~kBq  $^{241}$Am ($E_\alpha = 5.48$~MeV) spot is electro-deposited on a 6 mm diameter surface: we mask the active surface to reduce the  rate of triggering events to about 1 kBq. In liquid argon, alphas stop within few micrometers, therefore the scintillation source can be considered point-like and located at the source surface. $\alpha$ particles emitted close to the mask center produce the maximum light-cone and populate the peak structure of the energy spectrum, those emitted in proximity of the mask edges produce shadowed light cones and are responsible for the low energy tail,  as shown in Figure~\ref{fig:alpha_spectra}. The $\alpha$ full energy peak position is retrieved by fitting with a gaussian convolved with an exponential~\cite{PDE_XA_JINST}.\\
    The geometrical acceptance of the system is computed via a Montecarlo simulation considering only the maximum light-cone: the reflectivity of the 128~nm light steel cylindrical walls and parts is not considered since it has been measured to be negligible.\\

    \subsection{Photon Detection Efficiency}
    The Photon Detection Efficiency (PDE) is given by the ratio between the detected photons and the photons hitting the device entrance windows:
    \begin{equation}
        PDE = \frac{Q_\alpha}{Q_{sphe} f_{xt} f_{int-pur}} \cdot \frac{4\pi}{\Omega E_\alpha LY Q_f}\label{eq:PDE}
    \end{equation}
    The first term of Equation~\ref{eq:PDE} gives the number of detected photons: $Q_\alpha$ and $Q_{sphe}$ are the mean charge of the alpha full peak and of the single photoelectron peak respectively,  while $f_{xt}$ and $f_{int-pur}$, are the correction factors related to cross talk  and charge losses respectively. The latter is related to the limited integration window and to the actual quenching impurities concentration in the LAr; being anti-correlated in a way that their product remains almost constant (see \cite{Álvarez-Garrote2024} for more details), these are combined in one factor. The second term of Equation~\ref{eq:PDE} gives the number of scintillation photons reaching the entrance window. It includes the $\alpha$ particle energy $E_\alpha$, the liquid argon light yield $LY$, the $\alpha$ particle scintillation quenching factor $Q_f$ and the geometrical acceptance $\Omega/4\pi$. The values of the constant parameters (reported in Table \ref{tab:PDEconst}) are taken from literature \cite{Doke_2002,Hitachi1983,19-T01007_HPK_SiPMsCharacterizationCT}, while the variables are retrieved as described in the previous sections. The geometrical acceptance is computed for each source position by Montecarlo simulation as described in Section \ref{sec:MC}.

    \begin{table}[H]
        \centering
        \begin{tabular}{lc}
        \toprule
        Parameter   & Value\\
        \midrule
        $E_\alpha $ & 5.48 MeV \\
        Light Yield & 50'000 ph/MeV \\
        $Q_f$       & 0.70 \\
        $f_{xt}$ at 45\% PDE & 1.11 \\
        \bottomrule
        \end{tabular}
        \caption{PDE computation constants.}
        \label{tab:PDEconst}
    \end{table}

    \subsection{Geant4 based optical simulation}\label{sec:MC}
    A Montecarlo simulation developed with Geant4 \cite{GEANT4:2002zbu} was run to assess the impact on the PDE of the different X-Arapuca components and to explore what modifications to the baseline assembly resulted in a meaningful PDE improvement, to be later tested with laboratory measurements.\\
    The entire process is simulated: from alpha particles causing LAr scintillation to the photons being detected by the SiPMs, with the two stages of downshifting and the various reflective surfaces. The geometry is recreated in a simplified way, but with all the relevant characteristics; the alpha source is modeled with its holder in order to take into account shadowing effects. The materials optical properties are defined, at the best of our knowledge, with values from internal measurements (i.e. Sec~\ref{sec:Pavia}) and the literature.\\
    The module efficiency is given by the ratio between the detected photons and the scintillation photons hitting the optical windows. For each run, 100 alpha particles are generated in order to have adequate and consistent statistical errors. The alpha source can be moved along the module length in order to test uniformity, replicating the laboratory measurement procedure.\\
    The absolute efficiency obtained via the simulation is a factor $\sim$2.1 lower than the one measured in experimental tests. This discrepancy is attributed to some material properties not being precisely determined (and thus not being correctly simulated), such as PMMA mean absorption length and pTP quantum efficiency (or light yield) at cryogenic temperature. Nevertheless, relative differences obtained varying single parameters were proven to be consistent with experimental data, effectively guiding the optimization of the X-Arapuca.\\
\section{Measurements and results\label{data}}
The test setup demonstrated excellent  reproducibility between different runs, as shown in Figure~\ref{fig:HG-dimples} (top). Statistical errors are negligible, thanks to the high statistics of the alpha spectra. 
Systematics dominate the overall measurement error on the absolute PDE value but these are common to all the tested configurations, unless otherwise stated, so that relative differences are significant even if comparable with the reported errors. The main contributor to systematic uncertainty is the geometrical acceptance of the module for the incoming scintillation light ($\sim$7\%): the computation is done via the Geant4 simulation and strongly depends on the source mask dimensions and position in the chamber, which determines the light cone illuminating the module. The second topmost contributor is the uncertainty on the SiPMs gain ($\sim$5\%), followed by the uncertainty on the liquid argon purity ($\sim$2\%). The resulting total systematic error is $\sim$8.5\%.\\

\subsection{WLS-LG with dimpled edges}
\begin{figure}
    \centering
    \begin{subfigure}{0.5\textwidth}
        \includegraphics[width=1\linewidth] {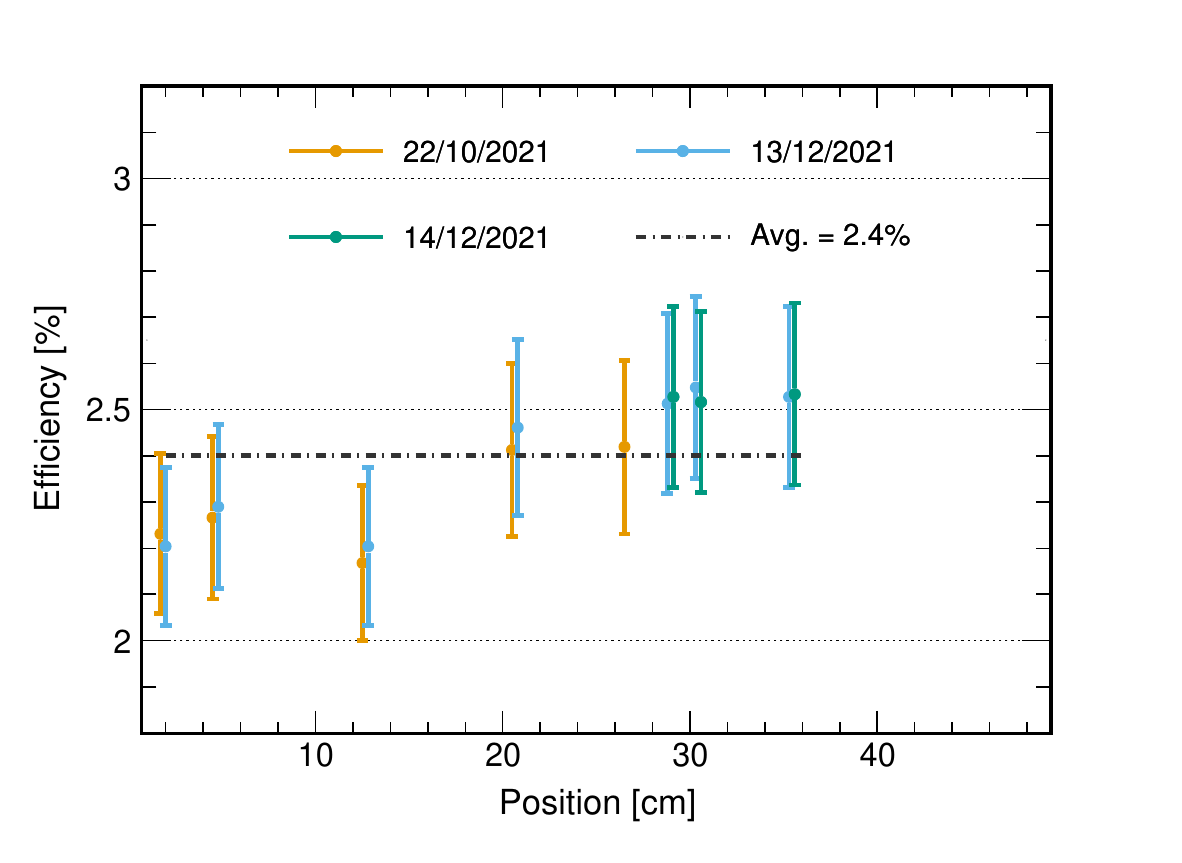} 
    \end{subfigure}\\
    \begin{subfigure}{0.5\textwidth}
        \includegraphics[width=1\linewidth]{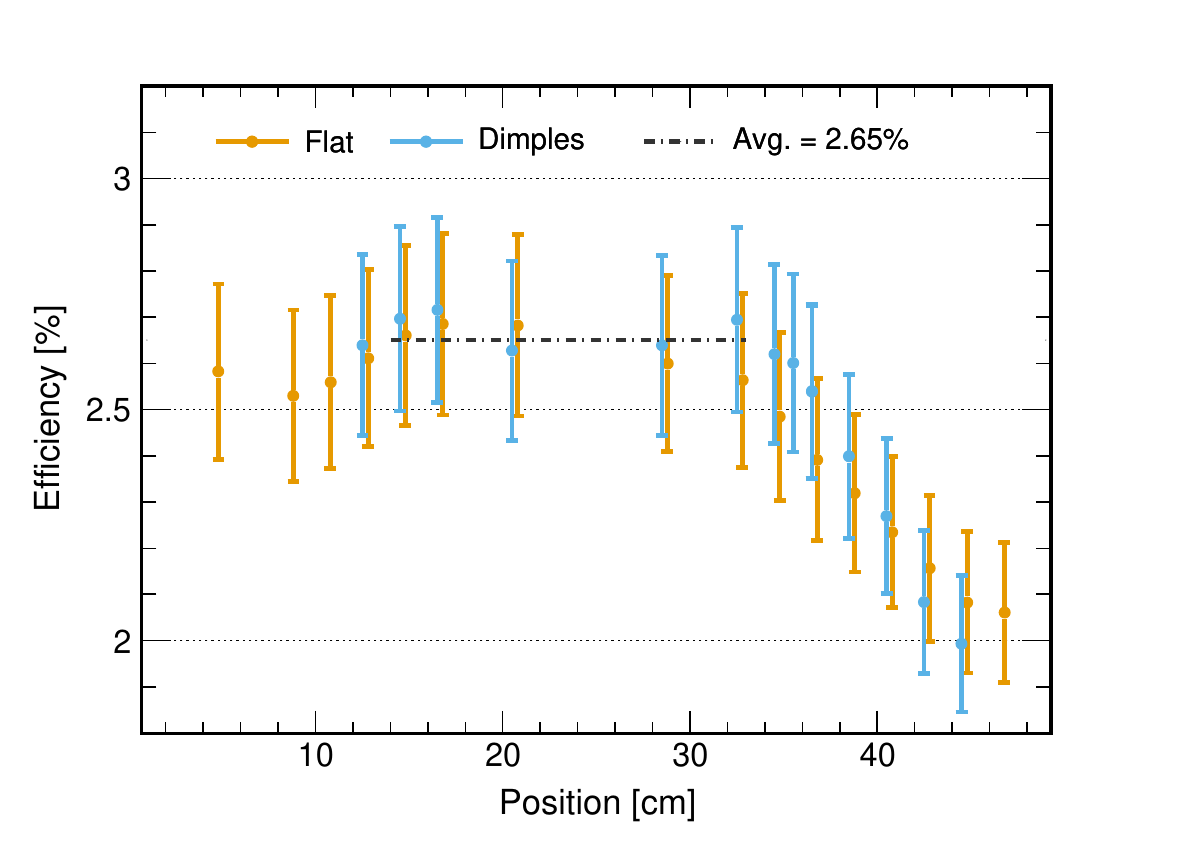}
    \end{subfigure}
    \caption{Top: The PDE of the baseline X-Arapuca \cite{Álvarez-Garrote2024}. Bottom: the PDE  of the modified module embedding two half-size WLS-LG, with (light-blue) and without (orange) edge dimples, as measured in this work.}
    \label{fig:HG-dimples}
\end{figure}

Reports from the collaboration \cite{Steklain_2023} indicated that dimples in the light guide, in front of the SiPMs, could help with light focusing and extraction towards the SiPMs, increasing the overall efficiency by 13\% and 17\% for square and cylindrical dimples respectively. Both cylindrical and square dimples have been tested in the MiB setup. In the standard XArapuca, the SiPMs are mounted on rigid circuit boards fixed to the frame, this would make difficult to achieve a consistent alignment with the dimpled WLS-LG. Therefore, for this measurement, in order to keep the SiPMs in contact with the dimpled WLS-LG, an X-Arapuca frame was modified in order to accept SiPMs mounted on spring-loaded kapton flex circuit boards.\\
This modified X-Arapuca frame was first equipped and tested with a flat light guide, reporting a PDE higher than the baseline X-Arapuca equipped with the same light guide (see Fig.~\ref{fig:HG-dimples}). This difference is mostly attributed to the improved optical contact at cold, as explained in Section~\ref{ols}.\\
The square and cylindrical dimples were then tested in a single LAr filling by equipping the modified X-Arapuca with two half-size light guides (240~mm $\times$ 92~mm), one with the long edges grooved with square dimples (8~mm wide, 1.2~mm deep, placed on top of the module), the other with cylindrical dimples (with a  4~mm radius, on the bottom of the module). The light guides were optically separated by applying Vikuiti reflector on the adjacent short edges at the center.\\
The bottom plot in Figure~\ref{fig:HG-dimples} shows the measured efficiency along the module length for the configurations with dimples (orange, cylindrical $<$25~cm, square$>$ 25~cm) and without dimples (blue). No significant difference is found between these three light guides. The drop in efficiency close to the top of the module was caused by a damaged pTP deposition on the uppermost dichroic filter and is therefore excluded from the overall efficiency computation.
The discrepancy between the simulations and the measurement was attributed to the fact that the dimple surface is more diffusive than the flat edge, as it is not possible to polish it after being laser cut. This, paired with the increased distance from the SiPMs in the cylindrical option, increases the amount of light escaping the light guide without impinging onto the SiPMs.

\subsection{WLS-LG optical sealing\label{ols}}
\begin{figure}
    \centering
    \begin{subfigure}{0.23\textwidth}
        \includegraphics[width=1\linewidth]{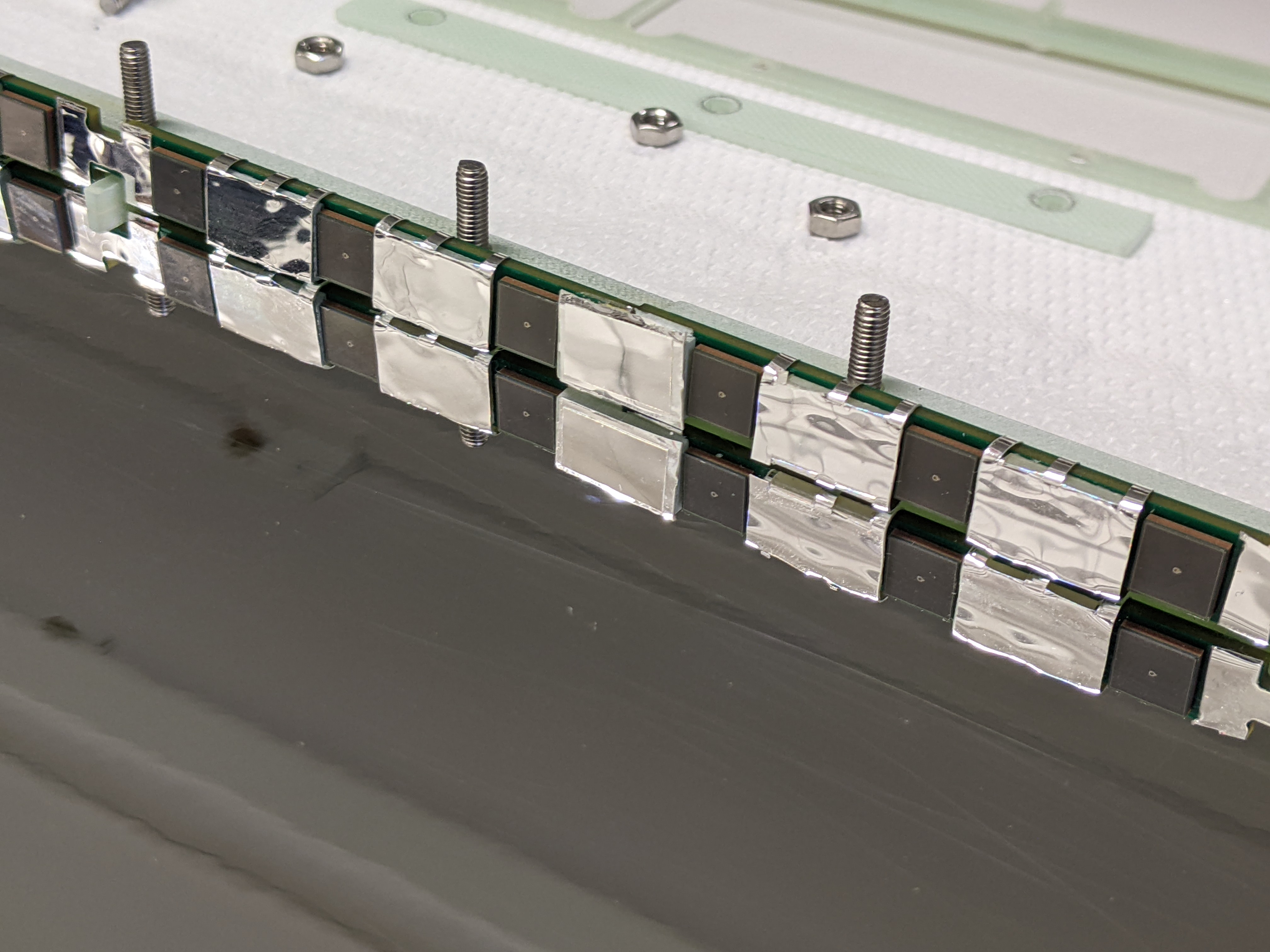}
    \end{subfigure}
    \begin{subfigure}{0.23\textwidth}
        \includegraphics[width=1\linewidth]{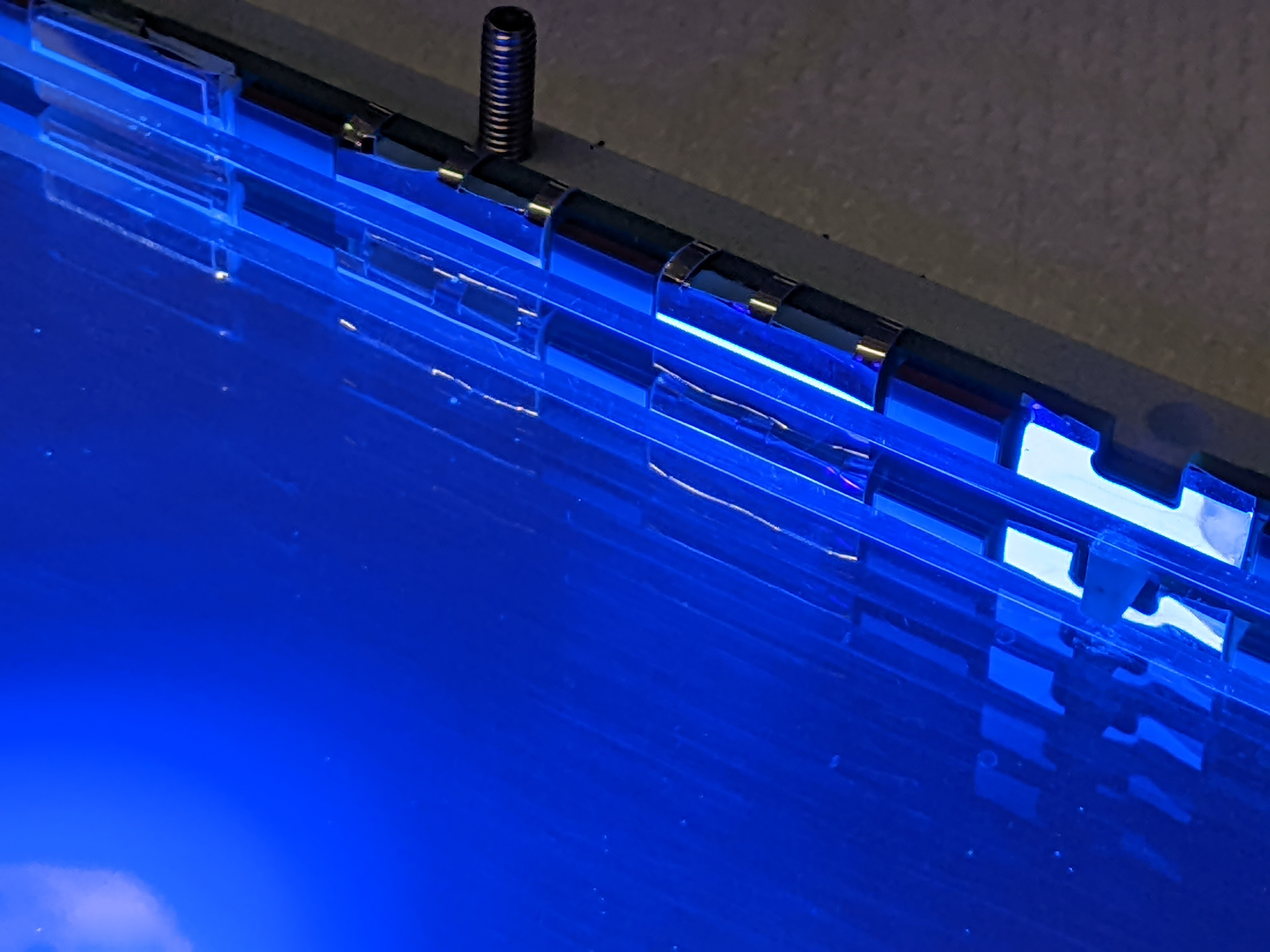}
    \end{subfigure}
    \caption{Pictures of the SiPM equipped sides of the module with Vikuiti reflector placed between the SiPMs with (right) and without (left) WLS-LG mounted. Brighter spots show where light is escaping.}
    \label{fig:VB}
\end{figure}

\begin{figure}
    \centering
    \begin{subfigure}{0.5\textwidth}
        \includegraphics[width=1\linewidth]{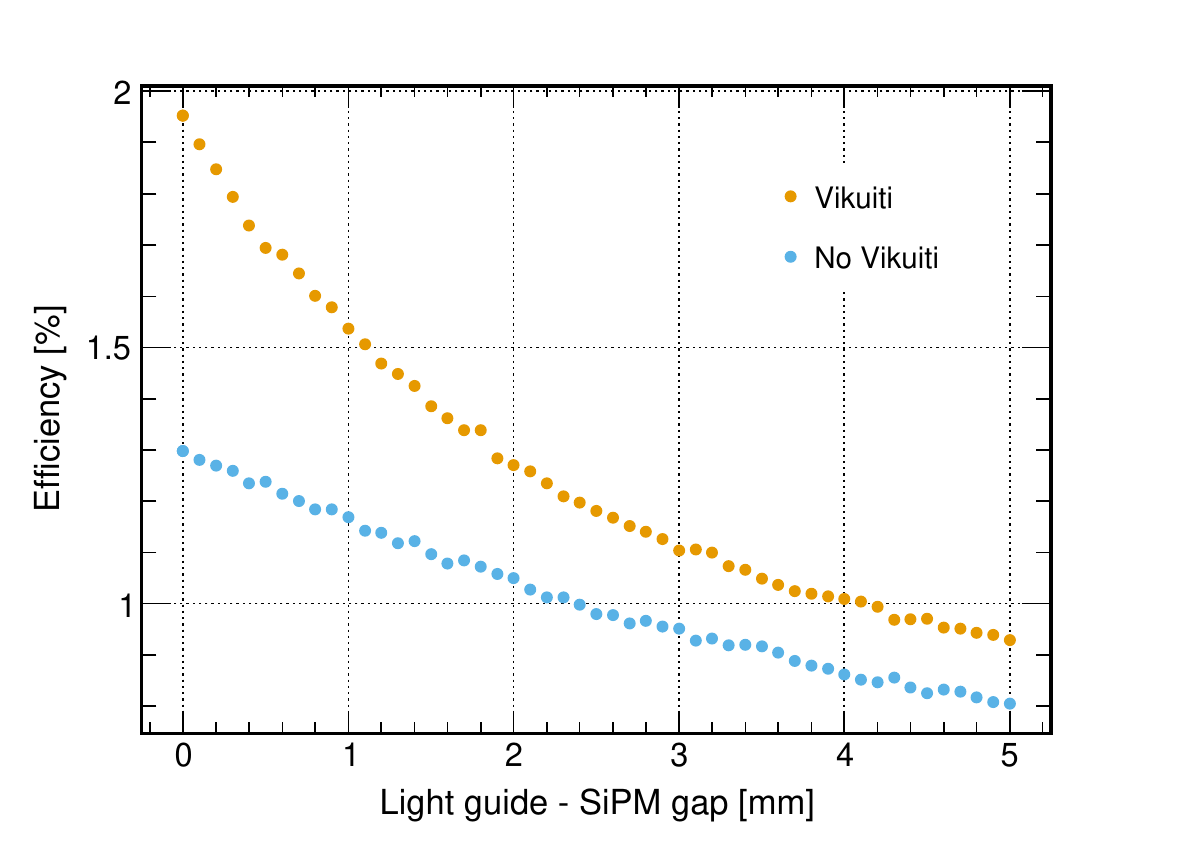}
    \end{subfigure}
    \begin{subfigure}{0.5\textwidth}
        \includegraphics[width=1\linewidth]{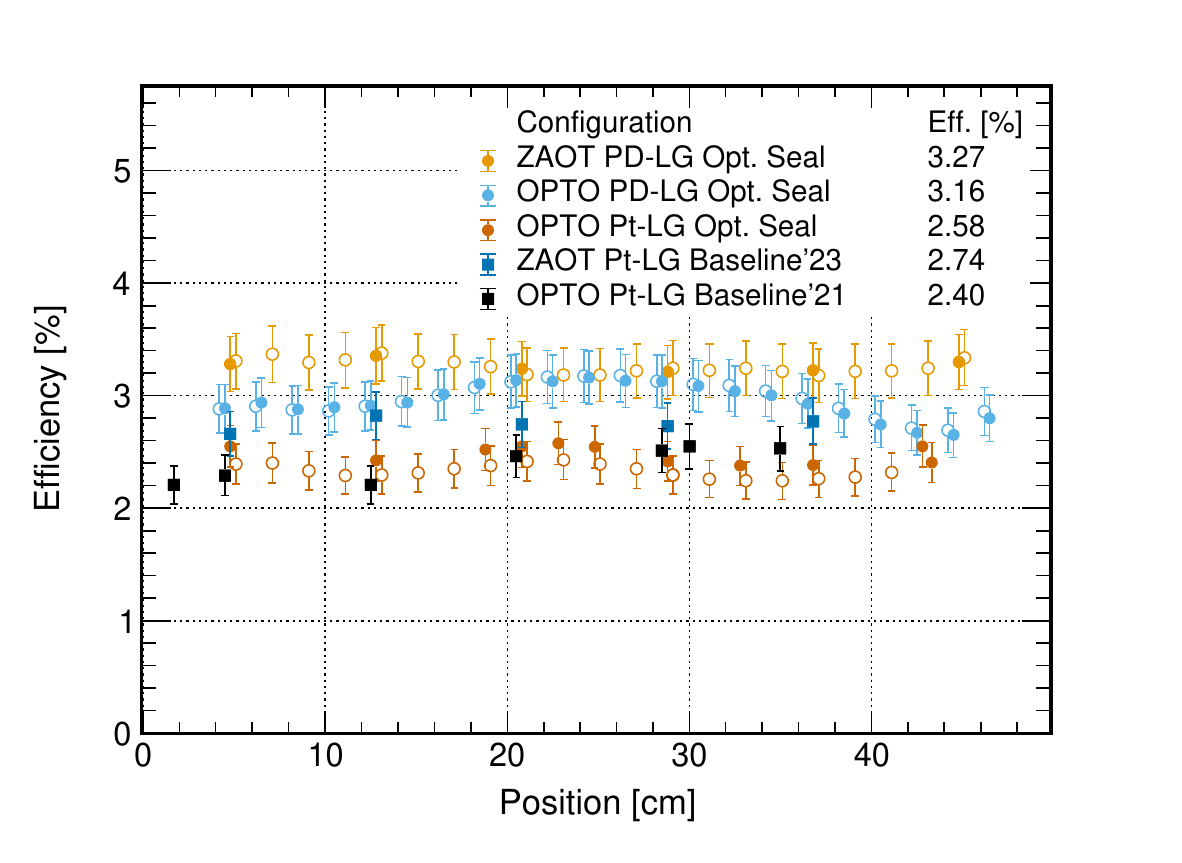}
    \end{subfigure}\\
    \begin{tabular}{r|cc|c}
        \toprule
        Configuration & ProtoDUNE & Prototype & WLS-LG \\
         \midrule
        Opt. Sealing & 3.27\% & 2.58\% & +27\% \\
        Baseline & 2.74\% & 2.40\% & +14\% \\
        \midrule
        Frame & +20\%  & +7.5\% & \\
        \bottomrule
    \end{tabular}
    \caption{Top: Montecarlo simulation for the configurations with (orange) and without (blue) Vikuiti applied closer to the light guide edge, varying the gap between SiPMs and light guide edge. Bottom: measurements of the module with Vikuiti reflector placed between SiPMs, hollow markers show the measurements performed the previous day. Measurement labeled "Pt-LG" utilized a prototype light guide, narrower than the other two measurements (PD-LG). The table shows relative improvements comparing different configurations.}
    \label{fig:VB_meas}
\end{figure}

In the baseline assembly of the X-Arapuca, photons can escape the light guide in the space between one SiPM and the next (as shown in Fig. \ref{fig:VB}), as the Vikuiti reflector is placed at a distance from the light guide that is equal or greater than the SiPM thickness. The escaped photons leave the module through the entrance windows or are absorbed by the X-Arapuca structural materials.\\
Simulations indicate that closing the gap between the WLS-LG and the SiPMs and reflector, improves the PDE by $\sim$40\% in the most realistic case (0.5~mm gap) and up to $\sim$50\% in the ideal case with no gap at cold temperature (see Fig.~\ref{fig:VB_meas} top).\\
To implement the optical sealing on the baseline assembly of the X-Arapuca, the option of applying Vikuiti on the long edges of the WLS-LG between the SiPMs was considered but discarded, as the different thermal contractions of the materials could potentially lead to a misalignment between the light guide and the X-Arapuca frame, resulting in Vikuiti partially obscuring the SiPMs. We then chose to place G10 rectangular pieces in the spaces between the SiPMs and to cover those with Vikuiti. The G10 pieces available for this test were 0.1~mm thicker than the SiPMs (Fig.~\ref{fig:VB}).\\
Being the SiPMs fixed to the frame, the PDE improvement in our tests is mainly driven by the light guide width, which determines the distance WLS-LG to SiPMs (Fig.~\ref{fig:VB_meas} bottom).
We performed tests equipping the module with a standard light guide ($\sim$93~mm wide, dimension used for all the light guides in ProtoDUNE-HD) and the WLS-LG prototype used for the previous paper measurement campaign \cite{Álvarez-Garrote2024} which is narrower, with a width of $\sim$92~mm. The 93~mm WLS-LG fits the X-Arapuca mechanical assembly at room temperature, leading to an estimated gap at LAr temperature of 0.5~mm per side between the guide and the sides covered by SiPMs and reflective blocks. The old WLS-LG has a $\sim$0.5~mm gap at room temperature and an estimated $\sim$1~mm gap at LAr per side. The measured configurations also differ for the equipped entrance windows, OPTO and ZAOT. The measurements are repotrted in Figure~\ref{fig:VB_meas} bottom.\\
The measured PDE of the device is 3.27\% with ZAOT DFs and 3.13\% with OPTO DFs. The overall efficiencies between the OPTO and ZAOT configurations are consistent; minor differences at 10~cm and 40~cm can be related to the pTP shifter degradation on the OPTO entrance window at the corresponding positions. No degradation of the pTP deposit on the ZAOT filter set was observed, hence the more uniform performance. The efficiency at the topmost positions is artificially increased  by the reflection of photons at the liquid argon surface.\\
In the same plot, we report a baseline measurement performed with the module mounting the protoDUNE WLS-LG, achieving an efficiency of 2.74\% and two measurements performed with the module mounting the prototype WLS-LG: the baseline from the previous paper (2.4\% PDE) and one with the module modified adding the Vikuiti reflector on G10 blocks (2.58\% PDE).
These results show a $\sim$20\% improvement over the baseline when comparing modules mounting the ProtoDUNE WLS-LG, lower than the value predicted by simulation. A lower improvement ($\sim$7.5\%) is measured with the prototype WLS-LG. While the lower improvement is expected for the narrower WLS-LG, the baseline measurement with the wider ProtoDUNE WLS-LG outperforms the module equipped with the narrower prototype WLS-LG and improved optical sealing, contradicting the simulation results shown in Figure~\ref{fig:VB_meas} top, as the width difference is not enough to cause a gap that would give such a result.\\
This discrepancy can be attributed to the non complete sealing of the WLS-LG edges: while in the simulation geometry, the full perimeter has been sealed (both for simplicity and to study the effect in an ideal case), in the measured device there are several sections that we could not seal. At the center of each of the eight SiPM boards, the WLS-LG supports prevented us to place the G10 blocks, providing escape points for the trapped light. This is clearly visible in Figure $\ref{fig:VB}$, on the right, where a bright spot can be seen in proximity of the light guide supports. Another possible cause can be the non uniform application of the Vikuiti reflector on the G10 blocks causing thickness differences between them.\\

\subsection{WLS-LG photon extraction}
\begin{figure}
    \centering
    \begin{subfigure}{0.5\textwidth}
        \includegraphics[width=1\linewidth]{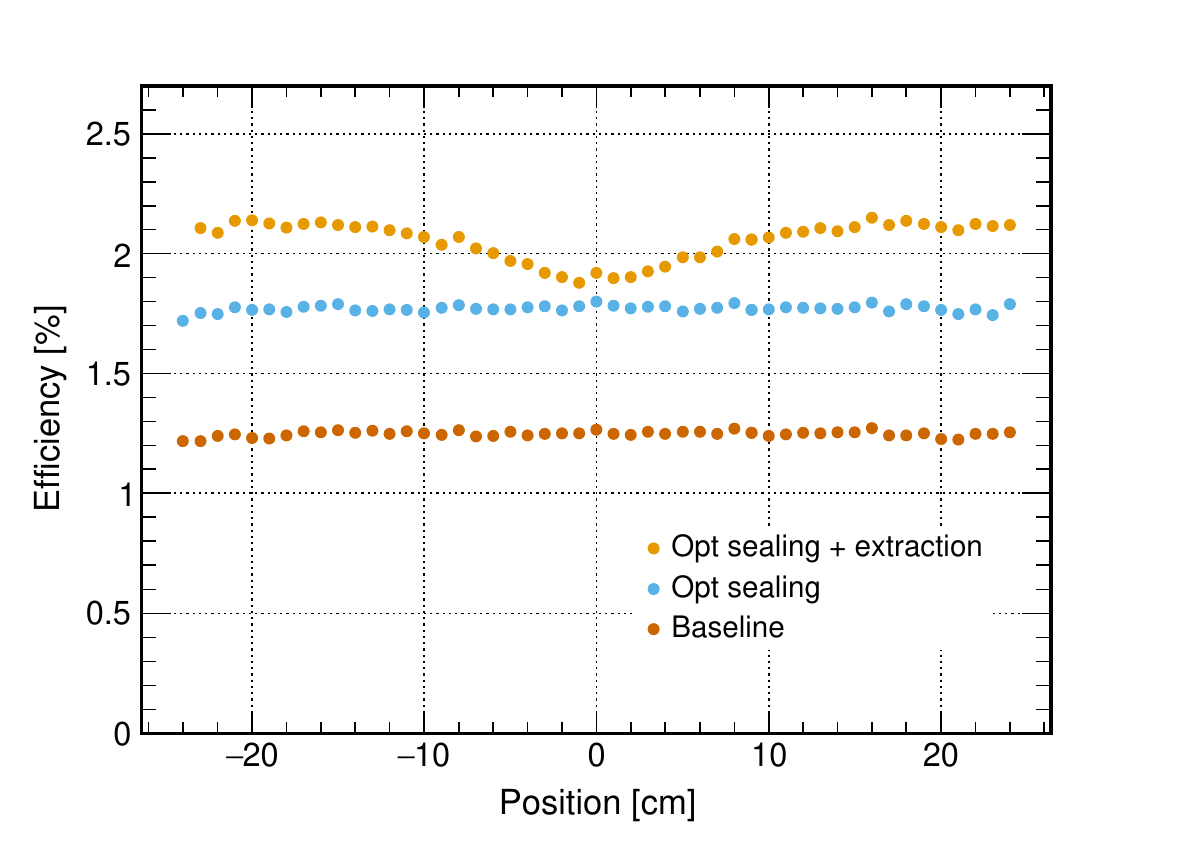}
    \end{subfigure}\\
    \begin{subfigure}{0.5\textwidth}
        \includegraphics[width=1\linewidth]{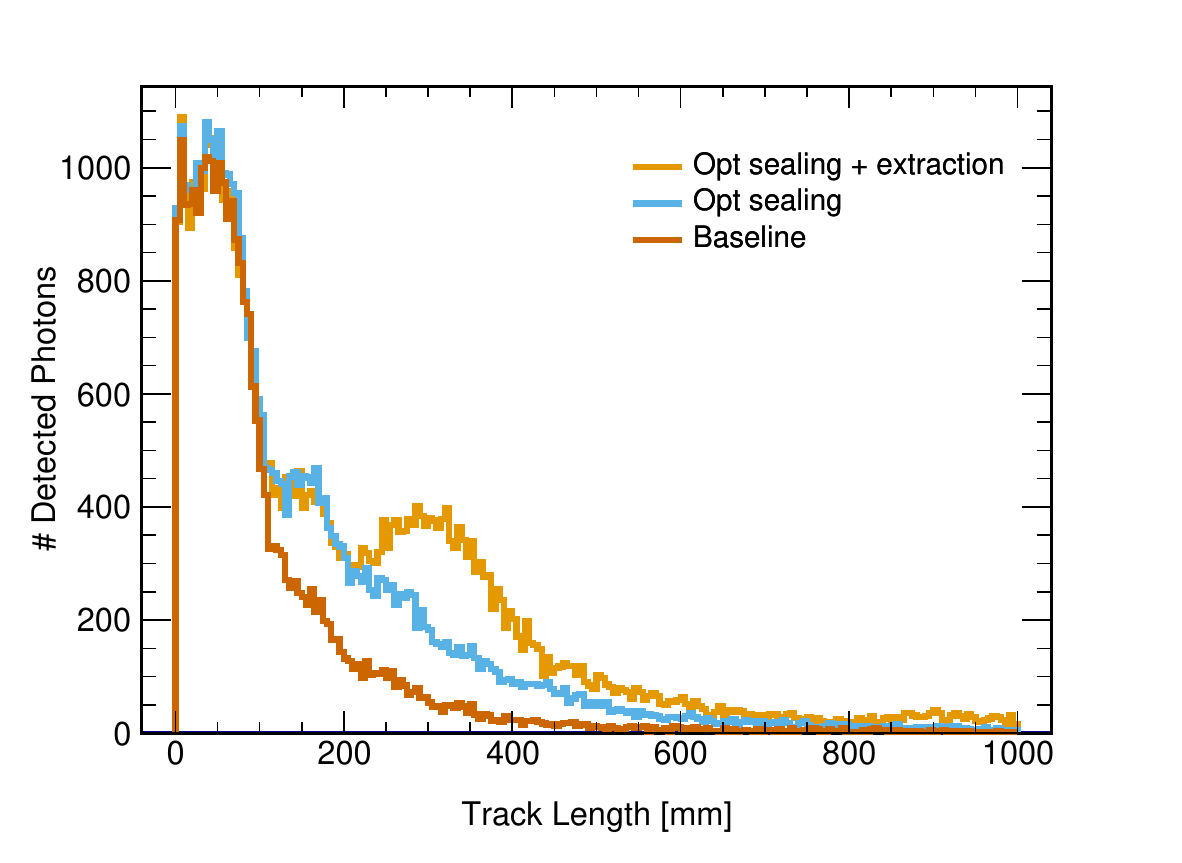}
    \end{subfigure}
    \begin{subfigure}{0.5\textwidth}
        \centering
        \resizebox{0.9\textwidth}{!}{%
        \begin{tikzpicture}

\definecolor{Blue}{RGB}{0,114,178}
\definecolor{SkyBlue}{RGB}{86,180,233}
\definecolor{Vermilion}{RGB}{213,94,0}
\definecolor{Orange}{RGB}{230,159,0}
\definecolor{bluishGreen}{RGB}{0,158,115}

\def\cutD{{4.16,0}}
\def\cutU{{5.84,2}}
\def\cutInc{{4.783251, 0.741688}}

\fill[blue!10] (0,0) rectangle (6.5,2); 
    \draw[thick] (0, 0) -- (0, 2);
    \draw[thick] (0, 0) -- (6, 0);
    \draw[thick] (0, 2) -- (6, 2);
    \draw[dashed, thick] (6, 2) -- (6.6, 2);
    \draw[dashed, thick] (6, 0) -- (6.6, 0);
\node at (5.6, 0.25) {Light guide};
\node at (6.1, -0.3) {LAr};
\draw[] (0.05,2.05) -- (6,2.05);
\draw[dashed] (6,2.05) -- (6.6,2.05);
\node at (2.5,-0.3) {SiPM side};
\draw[] (0.05,-0.05) -- (6,-0.05);
\draw[dashed] (6,-0.05) -- (6.6,-0.05);
\node at (2.5,2.3) {SiPM side};
\node[rotate=90] at (-0.3,1) {Vikuiti};
\node[rotate=50] at (5.6,1.3) {Vikuiti};
\draw[thick] (\cutD) -- (\cutU);

\def\origin{{0.1, 1.}}

\def\pcR{{4.793597, 0.754019}}
\def\pcD{{4.505940, 2.000000}}

\def\psRa{{0.463970, 2.000000}}
\def\psRb{{1.191911, 0.000000}}
\def\psRc{{1.919851, 2.000000}}
\def\psRd{{2.647792, 0.000000}}

\draw[thick, Vermilion] (\origin) -- (\pcR);
\draw[->, thick, Vermilion] (\pcR) -- (\pcD);

\draw[->, thick, dashed, Blue] (\psRc) -- (2.2, 1.3);
\draw[->, thick, Blue] (\psRb) -- (\psRc);
\draw[->, thick, Blue] (\psRa) -- (\psRb);
\draw[->, thick] (\origin) -- (\psRa);

\small
\node at (0.9, 1.8) {1$\rm{^{st}}$};
\node at (1.6, 0.2) {2$\rm{^{nd}}$};
\node at (2.4, 1.8) {3$\rm{^{rd}}$};

\tiny
\node[rotate=70] at (1.6, 1.5) {$\sim$10cm};
\node[rotate=-2] at (3.4, 0.95) {$\sim$24cm};

\end{tikzpicture}
        }
    \end{subfigure}
    \caption{Top: plot showing the overall PDE as a function of the source position along the module. Middle: histograms showing the track length of the detected photons with the $^{241}Am$ source at position 24~cm. Bottom: Top view of the X-Arapuca showing examples of detected photon tracks; the detection chances for photons hitting the SiPM equipped sides correspond to the peaks of the histogram above.}
    \label{fig:TL}
\end{figure}
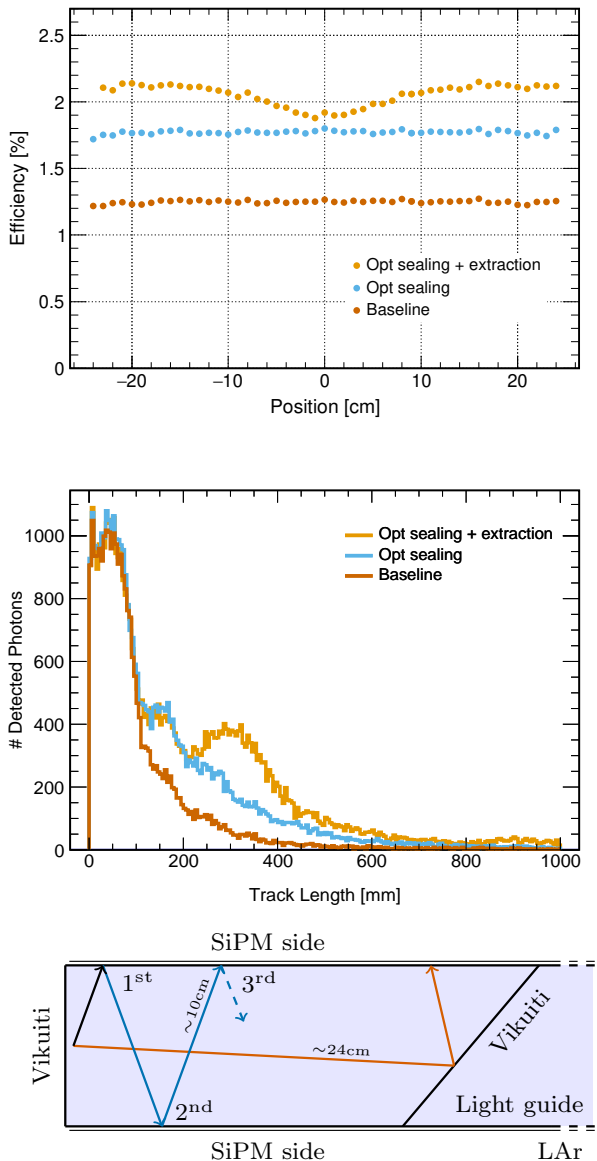

The problem that dimples would have solved is that of photons not being able to reach the SiPMs due to total internal reflection. 
The same boundary interaction exploited to propagate light towards the edges (in the vertical plane), also acts in the horizontal plane, so that photons emitted towards the module short edges cannot escape towards the SiPMs but stay trapped until they are absorbed, as showed in Figure~\ref{fig:tikz_XAcut}. 
The solution to this problem comes from a geometry that is able to recover the trapped photons while maintaining the polished light guide edge. This can be done cutting in half the light guide with an angle and lining the angled edges with Vikuiti. With this geometry, when the horizontally trapped photons hit the angled edge, they are reflected with a change of angle so that they can escape towards the SiPMs.
\begin{figure}
    \centering
    \resizebox{0.5\textwidth}{!}{%
    \begin{tikzpicture}

\def\cutD{{4.16,0}}
\def\cutU{{5.84,2}}
\fill[blue!10] (0,0) rectangle (10,2); 
\draw[thick] (0,0) rectangle (10,2); 
\draw[thick] (\cutD) -- (\cutU);
\node at (8.9, 1.7) {Light guide};
\node at (9.5, 2.3) {LAr};
\draw[] (0.05,2.05) -- (9.95,2.05);
\node at (5,-0.3) {SiPM side};
\draw[] (0.05,-0.05) -- (9.95,-0.05);
\node at (5,2.3) {SiPM side};
\node[rotate=90] at (-0.3,1) {Vikuiti};
\node[rotate=-90] at (10.3,1) {Vikuiti};

\def\origin{{6, 1.10}}
\def\pA{{10.000000, 0.300721}}
\def\pB{{8.495038, 0.000000}}
\def\pC{{4.783251, 0.741688}}
\def\pD{{5.072423, 0.000000}}
\def\pE{{0.000000, 1.697476}}
\def\pF{{1.513984, 2.000000}}

\draw[->, thick, color=green] (\origin) -- (\pA);
\draw[->, thick, color=green] (\pA) -- (\pB);
\draw[->, thick, color=green] (\pB) -- (\pC);
\draw[->, thick, color=green] (\pC) -- (\pD);

\draw[->, thick,dashed] (\pC) -- (\pE);
\draw[->, thick, dashed] (\pE) -- (\pF);
\draw[->, thick, dashed] (\pF) -- (\origin);

\end{tikzpicture}
    }
    \caption{Top view of the geometry of a WLS-LG divided in two pieces with a 40° cut in the middle. In green, the optical path of a photon if the reflective cut is present. The dashed line is the looping path in case the cut is not present.}
    \label{fig:tikz_XAcut}
\end{figure}
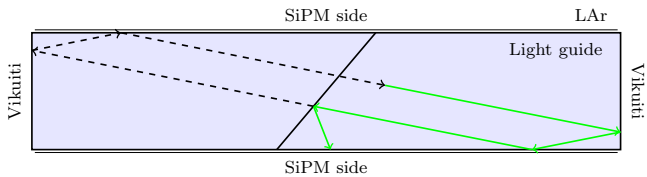%
\begin{figure}
    \centering
    \begin{subfigure}{0.22\textwidth}
        \includegraphics[width=1\linewidth]{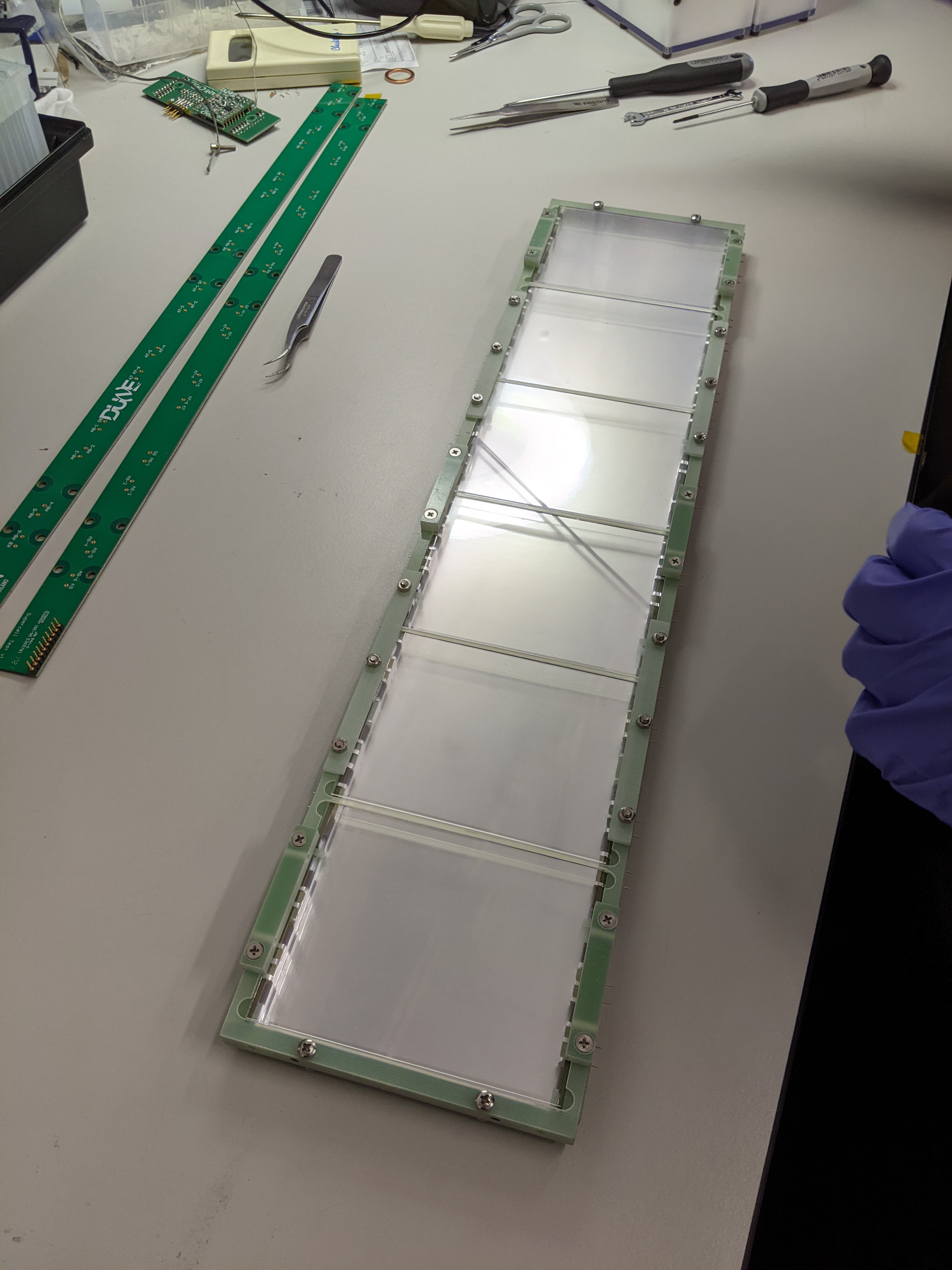}
    \end{subfigure}
    \begin{subfigure}{0.22\textwidth}
        \includegraphics[width=1\linewidth]{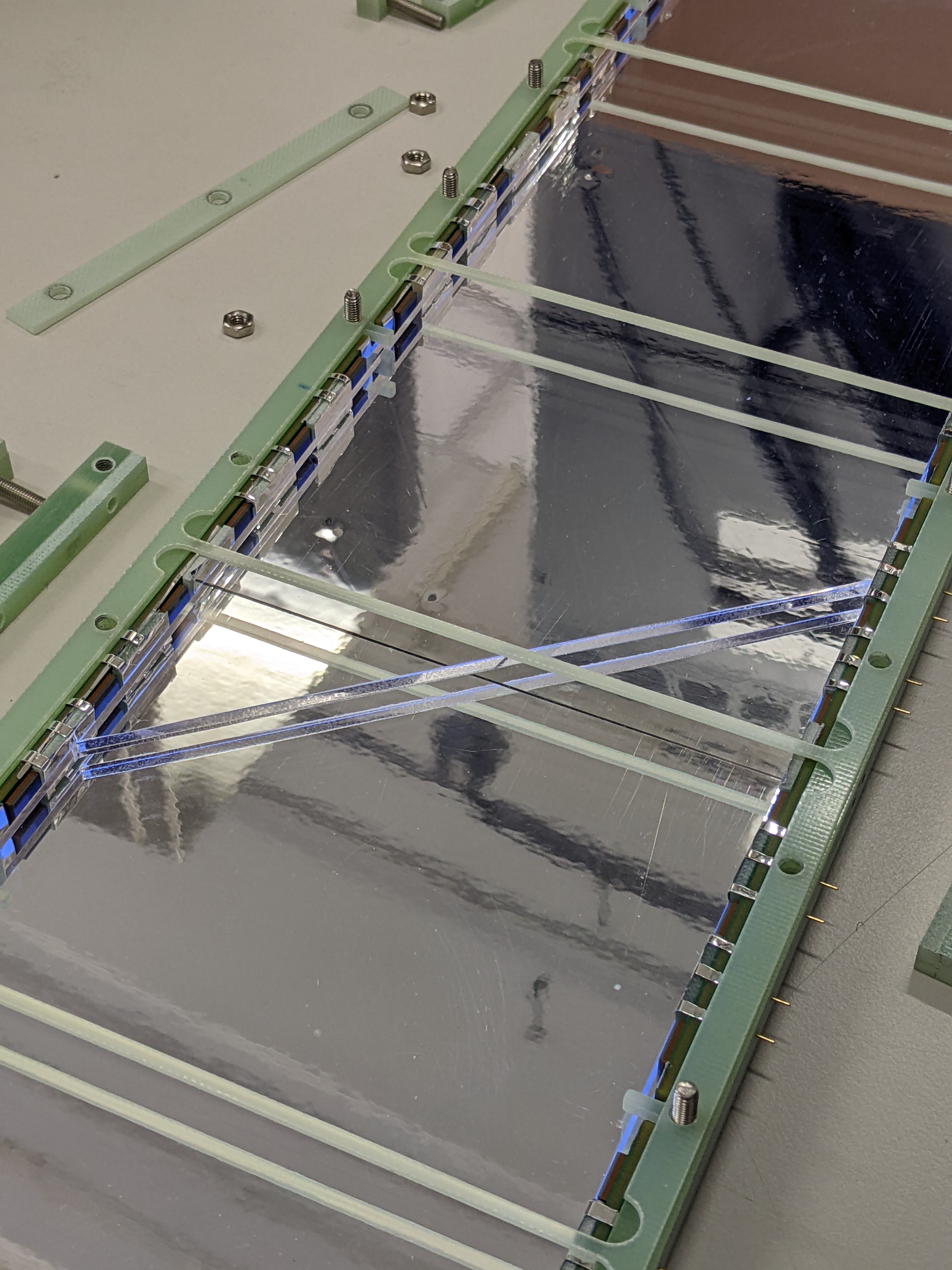}
    \end{subfigure}
    \caption{Picture of the module with 2 piece light guide. On the right without the entrance windows.}
    \label{fig:cut_pic}
\end{figure}
\begin{figure}
    \centering
    \begin{subfigure}{0.45\textwidth}
        \includegraphics[width=1\linewidth]{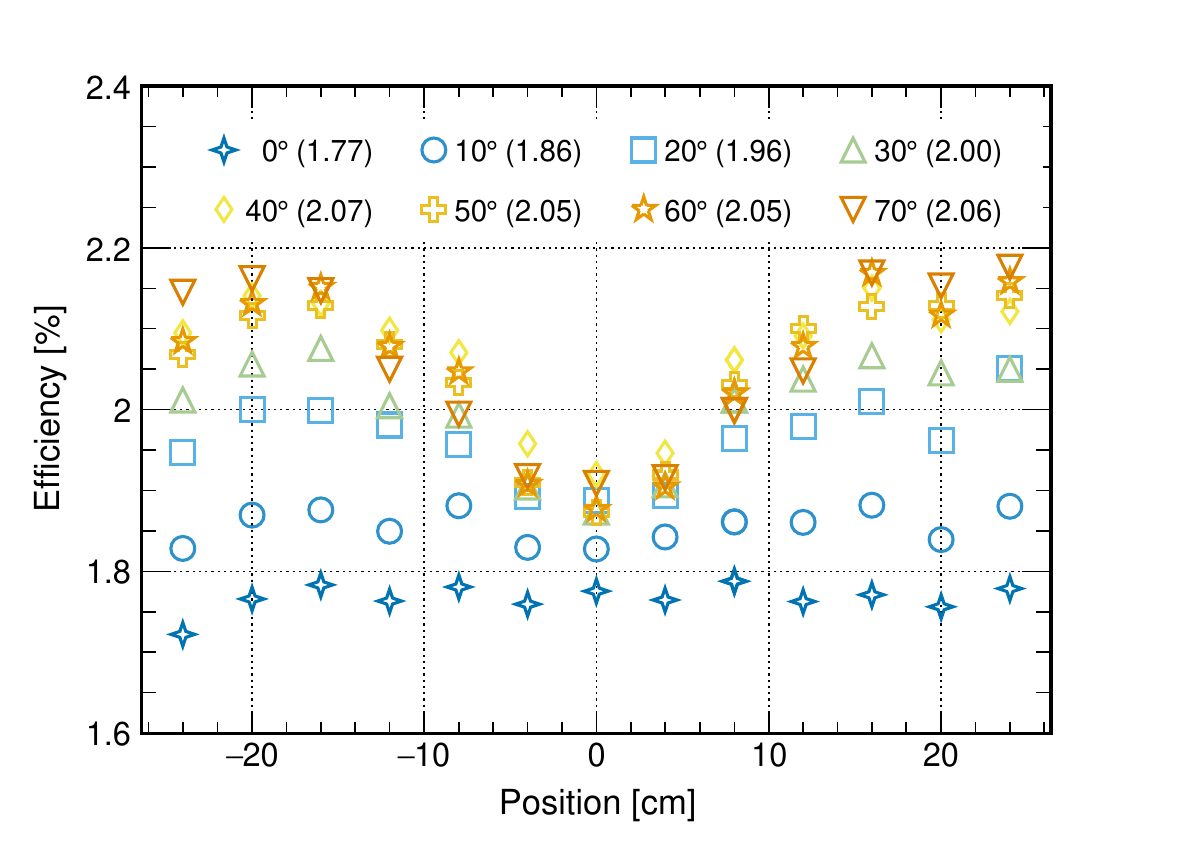}
    \end{subfigure}
    \caption{Source position dependence simulated for different cut angles. The average efficiency over all the positions is reported for each angle configuration.}
    \label{fig:ang_sim}
\end{figure}
Montecarlo simulations were performed by cutting the light guide in two or more pieces and with different angles, and the best compromise between PDE improvement, ease of construction and cost effectiveness was reached by cutting the light guide in two pieces; the angle of the cut in the middle was varied from 0° (cut parallel to the short edges), to 70° (Fig.~\ref{fig:ang_sim}), with 90° being parallel to the long edges, and the best result was given by a 40° angle cut, with an overall efficiency increase of $\sim$17\%.
The presence of the cut causes non-uniformity in the collection of photons along the length of the module, and is due to photons emitted around the center of the module, that would be immediately detected by close SiPMs if the cut would not reflect them away. The larger the angle, the higher this non-uniformity and, over a certain threshold, the PDE is negatively impacted.\\
Figure~\ref{fig:TL} top shows the simulation of three configurations: the baseline (green), better light guide light sealing (blue), with a $\sim$40\% PDE improvement over baseline, and the latter with better light sealing and the 40° cut light guide (orange), with a $\sim$17\% PDE improvement over the previous configuration. All other parameters of the geometry, crucially the light guide width, are the same for all the three configurations.
Figure~\ref{fig:TL} middle shows the track length of detected photons, from the emission point inside the light guide to the SiPM, in the case the source is positioned over one of the module ends ($\pm$24~cm in the top plot). The first peak, around $\sim$5~cm and common to all the configurations, represents photons that are detected upon the first reaching of a long edge equipped with SiPMs. The second peak, around $\sim$15~cm (and smeared with the next ones placed at increments of $\sim$10~cm), represents photons reflected upon the first reaching of an edge and detected at the opposite side of the module (and so on); this peak is enhanced for the last two configurations thanks to the better sealing of the light guide that allows more photons to be reflected to the opposite side rather than escape from the gaps between SiPMs. The last peak, observable at $\sim$30~cm in the cut light guide configuration, is given by photons emitted towards the short edges that reach the cut placed in the middle ($\sim$24~cm) and are reflected towards a long egde ($\sim$~5cm). A diagram showing examples of the possible detected photon tracks can be seen in Figure~\ref{fig:TL} bottom.\\
\begin{figure}[ht]
    \centering
    \begin{subfigure}{0.5\textwidth}
        \includegraphics[width=1\linewidth]{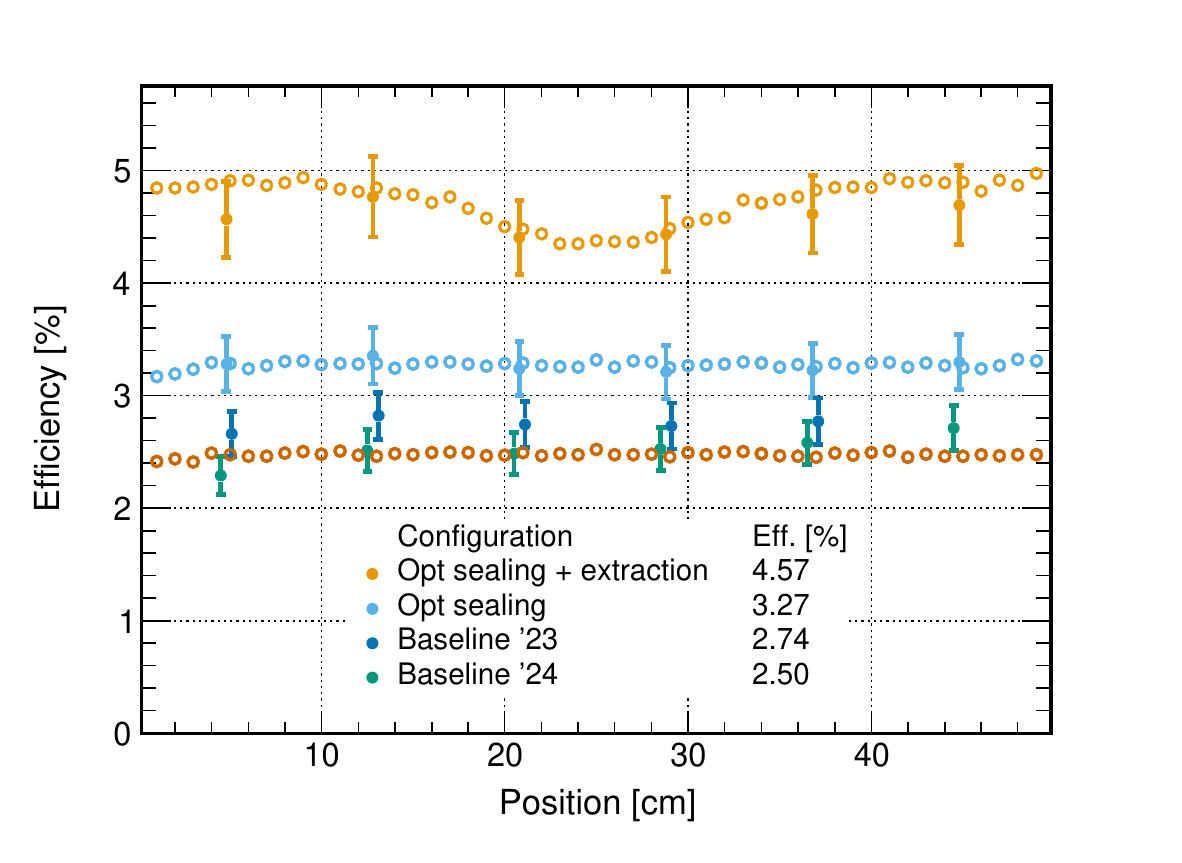}
    \end{subfigure}\\
    \begin{subfigure}{0.5\textwidth}
        \includegraphics[width=1\linewidth]{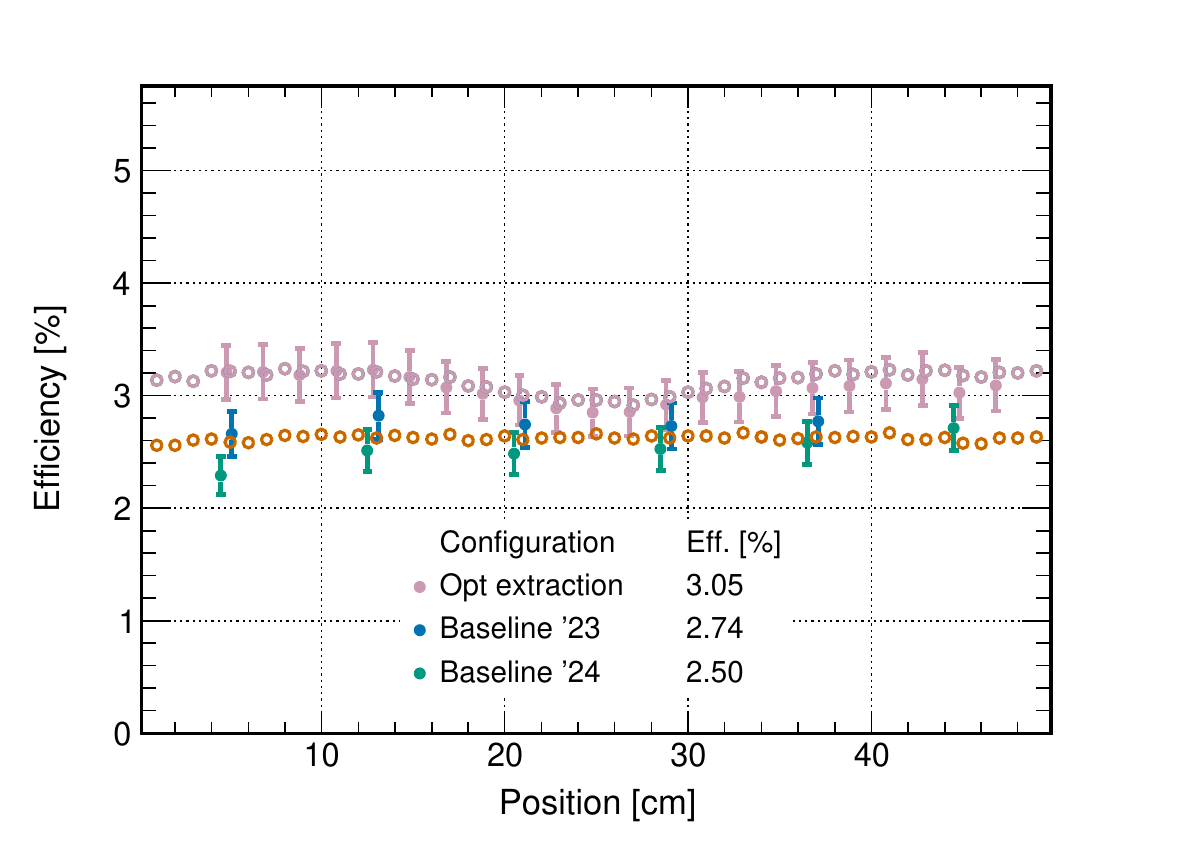}
    \end{subfigure}
    \caption{Measurements of the module with improved optical extraction versus the baseline, with (top) and without (bottom) improved optical sealing. The measurements are superimposed to the simulation outputs rescaled by a factor $\sim$2.1 and custom light guide widths.}
    \label{fig:cut_meas}
\end{figure}
A first measurement has been performed on a module with both the cut light guide and the improved light sealing as described in the previous section. The light guide has been cut with the 40° angle, polished and applied Vikuiti reflector on both of the newly cut edges, as can be seen from Figure~\ref{fig:cut_pic}. The overall efficiency obtained from the measurement is much higher than what was expected according to the simulation, with a measured efficiency improvement of $\sim$40\% over baseline. This can be explained by a greater width of the cut light guide, $\sim$93.5~mm, compared to the one utilized for the previous measurements, $\sim$93~mm. The same effect that negated the improvement of the increased optical sealing due to a narrower WLS-LG (see Fig.~\ref{fig:VB_meas}), is here acting in the opposite way, boosting the overall performance. In fact, the cut light guide is slightly wider than the internal module cavity, so that at room temperature the frame is flexing outwards; cooling down during LAr filling, the frame likely relaxes compensating the gap that would form due to the light guide thermal contraction. Looking at the top plot in Figure~\ref{fig:VB_meas}, it can be seen that for gaps $<$0.5~mm, that would be usually inaccessible, the efficiency dependence becomes stronger, meaning that for a gap approaching zero, this effect would provide a great improvement.\\
This measure, shown in the top plot of Figure~\ref{fig:cut_meas}, is therefore affected by two different effects, difficult to disentangle. To determine the efficiency improvement given by the cut light guide only, the same module has been measured without the Vikuiti reflector stands between SiPMs. As it can be seen from the same top plot in Figure~\ref{fig:VB_meas}, the configuration without such stands is less affected by the difference in gap between SiPMs and light guide, making the effect caused by the difference in light guide width subdominant compared to the effect of interest for the measurement. In the bottom plot of Figure~\ref{fig:cut_meas}, it is shown the measurement without the Vikuiti stands. It can be seen how the efficiency is lower than the previous measurements, and the improvement over the baseline of $\sim$22\%, more comparable with the one predicted via simulation.\\
All the reported measurements can be recreated within the Geant4 simulation, taking into account the 2.1 discrepancy factor on the efficiency absolute value and estimating the difference in width of the light guide for the different configurations. The simulation output in Figure~\ref{fig:cut_meas} has been obtained with a reduced gap between the light guide edge and SiPMs/Vikuiti of 0.1mm for the cut light guide, reproducing its greater width, while for the other configurations we assumed the nominal 92mm width.\\

\subsection{X-Arapuca entrance windows}
\label{sec:entrancewindow}
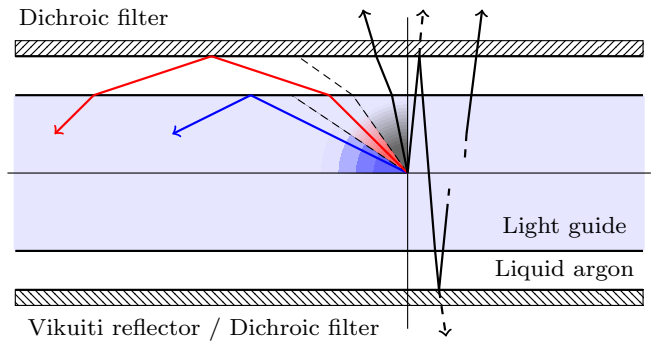
\begin{figure}
    \centering
    \resizebox{0.5\textwidth}{!}{%
    \begin{tikzpicture}

\fill [blue!10] (-5,1) rectangle (3,-1);

\pgfdeclareradialshading{sblack}{\pgfpoint{0cm}{0cm}}%
{color(0cm)=(black!60);
color(0.3cm)=(black!60);
color(0.8cm)=(blue!10);
color(10cm)=(blue!10)}
\begin{scope}
    \clip (0,0) rectangle (-5, 1);
    \fill[shading=sblack]  (-1,-1) rectangle (1,1);
\end{scope}
\pgfdeclareradialshading{sred}{\pgfpoint{0cm}{0cm}}%
{color(0cm)=(red!60);
color(0.15cm)=(red!60);
color(0.4cm)=(blue!10);
color(5cm)=(blue!10)}
\begin{scope}
    \clip (0,0) --  (0, 0) -- (-0.70,1) -- (-1.48,1) -- cycle;
    \fill[shading=sred]  (-2,-2) rectangle (2,2);
\end{scope}
\pgfdeclareradialshading{sblue}{\pgfpoint{0cm}{0cm}}%
{color(0cm)=(blue!60);
color(0.06cm)=(blue!60);
color(0.16cm)=(blue!10);
color(10cm)=(blue!10)}
\begin{scope}
    \clip (0,0) --  (0, 0)-- (-1.48,1) -- (-5, 1) -- (-5, 0) -- cycle;
    \fill[shading=sblue]  (-5,-5) rectangle (5,5);
\end{scope}

\draw[very thick, blue!10] (-5,0.99) -- (-5,-0.99);
\draw[very thick, blue!10] (3,0.99) -- (3,-0.99);

\draw[thin] (0,-2) -- (0,2);
\draw[thin] (-5.1,0) -- (3.1,0);

\draw[thick] (-5,1.5) -- (3,1.5);   \node at (-4, 2) {Dichroic filter};
    \draw[pattern=north east lines] (-5,1.5) rectangle (3,1.7);
                                    \node at (2, -1.25) {Liquid argon};
\draw[thick] (-5,1) -- (3,1);       \node at (2, -0.7) {Light guide};
\draw[thick] (-5,-1) -- (3,-1);
\draw[thick] (-5,-1.5) -- (3,-1.5);       \node at (-2.6, -2) {Vikuiti reflector / Dichroic filter};
    \draw[pattern=north west lines] (-5,-1.5) rectangle (3,-1.7);


\draw[densely dashed] (0,0) -- (-1.48,1);
\draw[densely dashed] (0,0) -- (-0.70,1);
\draw[densely dashed] (-0.70,1) -- (-1.40,1.5);




\draw[thick, blue] (0, 0) -- (-2, 1);
    \draw[->, thick, blue] (-2, 1) -- (-3, 0.5);

\draw[thick, red] (0, 0) -- (-1, 1);
    \draw[thick, red] (-1, 1) -- (-2.5, 1.5);
    \draw[thick, red] (-2.5, 1.5) -- (-4, 1);
    \draw[->, thick, red] (-4, 1) -- (-4.5, 0.5);

\draw[thick] (0, 0) -- (-0.2, 1);
    \draw[thick] (-0.2, 1) -- (-0.4, 1.5);
    \draw[thick] (-0.4, 1.5) -- (-0.45, 1.7);
    \draw[->, thick] (-0.45, 1.7) -- (-0.57, 2.1);

\draw[thick] (0, 0) -- (0.1, 1);
    \draw[thick] (0.1, 1) -- (0.15, 1.5);
        \draw[dashed, thick] (0.15, 1.5) -- (0.16, 1.7);
        \draw[->, dashed, thick] (0.16, 1.7) -- (0.2, 2.1);
    \draw[thick] (0.15, 1.5) -- (0.35, -1);
    \draw[thick] (0.35, -1) -- (0.40, -1.5);
    \draw[thick] (0.40, -1.5) -- (0.45, -1);
        \draw[->, dashed, thick] (0.40, -1.5) -- (0.49, -2.1);
    \draw[thick] (0.45, -1) -- (0.50, -0.5);
    \draw[thick, dashed] (0.50, -0.5) -- (0.53, -0.2);
    \draw[thick, dashed] (0.73, 0.2) -- (0.75, 0.5);
    \draw[->, thick] (0.75, 0.5) -- (0.95, 2.1);
    
\end{tikzpicture}
    }
    \caption{X-Arapuca section diagram showing the light components trapped by the light guide (blue, A) and by the dichroic filters (red, B) or that escapes the X-Arapuca (black).}
    \label{fig:trapping_components}
\end{figure}

\begin{figure}
    \centering
    \includegraphics[width=\linewidth]{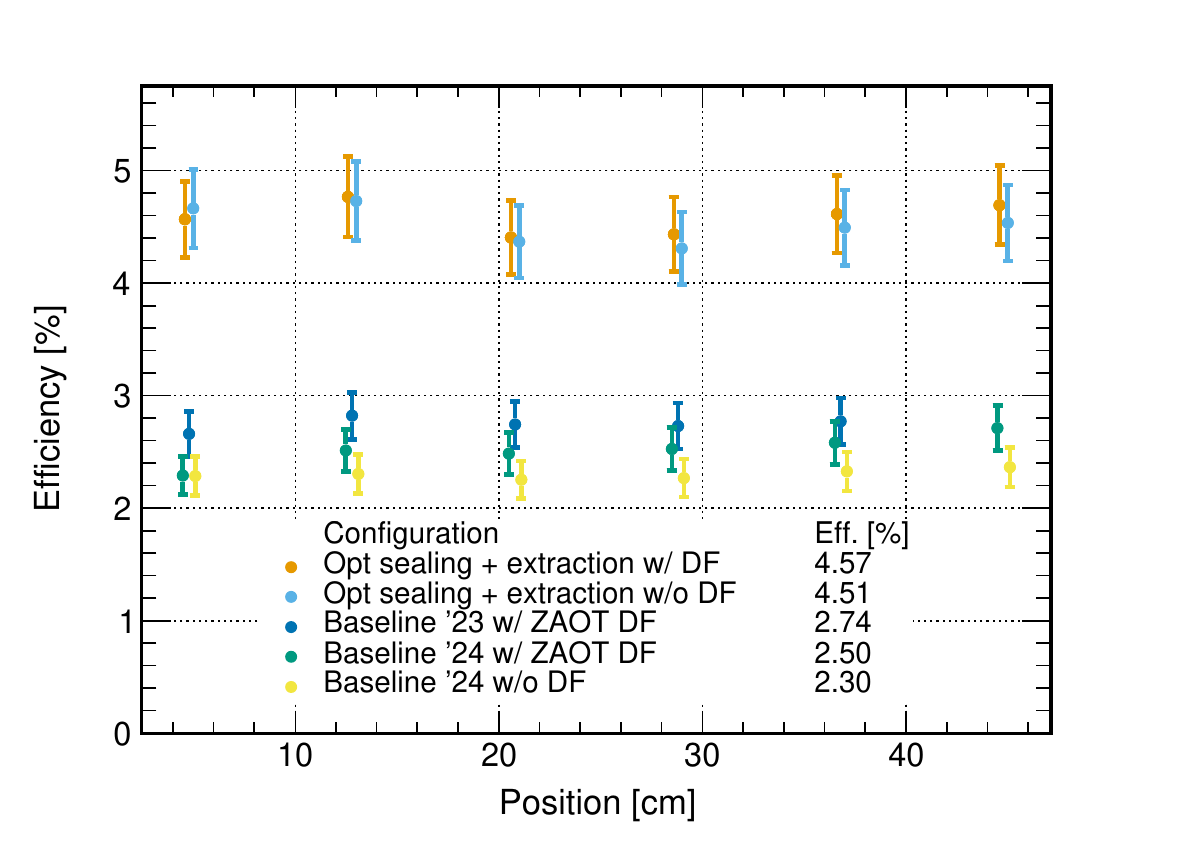}
    \caption{Efficiency measurement of SC configurations with and without dichroic filters for baseline and improved photon collection configurations.}
    \label{fig:noDF_meas}
\end{figure}

As discussed previously, and shown in Figure~\ref{fig:ZAOT_TC}, the dichroic filter coating is not fully transparent to pTP photons nor fully reflective to those escaping the WLS-LG. The dichroic transmittance of pTP photons affects both the WLS-LG and dichroic-coating trapping mechanisms, therefore, if the WLS-LG trapping becomes dominant, the presence of the dichroic filter causes a negative net effect.\\
The amount of photons coming from the first shifter, downshifted in the light guide and collected onto a SiPM can be approximately expressed by the formula:
\begin{equation}
    \gamma_{collected} = T \cdot ( A + B) \cdot \gamma_{pTP}
\end{equation}
where $\gamma_{pTP}$ are the 350~nm photons coming from pTP, $T$ is the transparency of the entrance window to these incoming photons, and $A$ and $B$ are the fraction of photons trapped by the light guide and dichroic filters respectively, as can be seen in blue and red in Figure~\ref{fig:trapping_components}.\\
At first approximation, considering Fresnel reflections at the surface, the transparency of the clear entrance window can be considered as T=0.96 while, with the dichroic deposition it can be assumed to be $T=0.70$ (from the measurement reported in Section~\ref{sec:Pavia}). The $A$ and $B$ components in this equation depend, in the first place, on the geometrical capture efficiency, constant at any point of the light guide. Given the critical angle $\theta_c$=56° over which total internal reflection (TIR) takes place, the fraction of downshifted photons trapped into the light guide due to TIR is $A=0.56$. The dichroic filters are designed to reflect light impinging with angles over 45° so, taking into account the refraction taking place at the light guide - argon interface, a fraction $B=0.25$ of the photons emitted in the light guide can be assumed to be trapped.
These two constants are to be considered as upper limits; the components non ideality, as well as the losses during the photon propagation towards the light guide edges can be modeled assigning an efficiency $\epsilon_A$ and $\epsilon_B$ to each one of the constants A and B:
\begin{equation}
    \gamma_{collected} = T \cdot ( 0.56\ \epsilon_A + 0.25\ \epsilon_B) \cdot \gamma_{pTP}
\end{equation}
These efficiencies depend on multiple variables such as module dimensions, surface finishes, optical couplings, etc. These variables affect the three main functions of the X-Arapuca: photon trapping, transport towards the edges and extraction/detection by the sensors. Some module properties affect the overall performance via the dependence on multiple variables. In example, if the module dimension increases, photons emitted further from the edges have to travel more to reach a sensor; the fraction of photons reaching the edges depends on the mean absorption length in the light guide (and/or LAr) and the probability of escaping at each surface reflection. The longer a photon has to travel and the more reflections have to take place (the latter also dependent on light guide thickness), the less photons will be collected at the edges. Depending on the X-Arapuca device properties, different effects can be dominant in determining the value of $\epsilon_A$ and $\epsilon_B$.\\
The light guide photon collection efficiency $\epsilon_A$ depends mostly on the PMMA transparency and on the extraction efficiency of photons towards the SiPMs, while transport via TIR is very efficient with a probability of a photon escaping due to surface imperfections $<$1\%.
The dichroic photon collection efficiency $\epsilon_B$ depends mostly on the quality of the dichroic filter and the dimensions of the module. Producing a dichroic filter capable of reflecting photons with a wide range of wavelenghts and incidence angles comes with the cost of non perfect reflectivity in the region of interest. Given the requirements of the X-Arapuca, it has been difficult obtaining reflectivities above 98\%, this implies that the mean number of possible reflections (thus distance traveled by photons) before a photon escapes is lower for dichroic filters trapping than for light guide trapping.\\
\begin{figure}
    \centering
    \includegraphics[width=\linewidth]{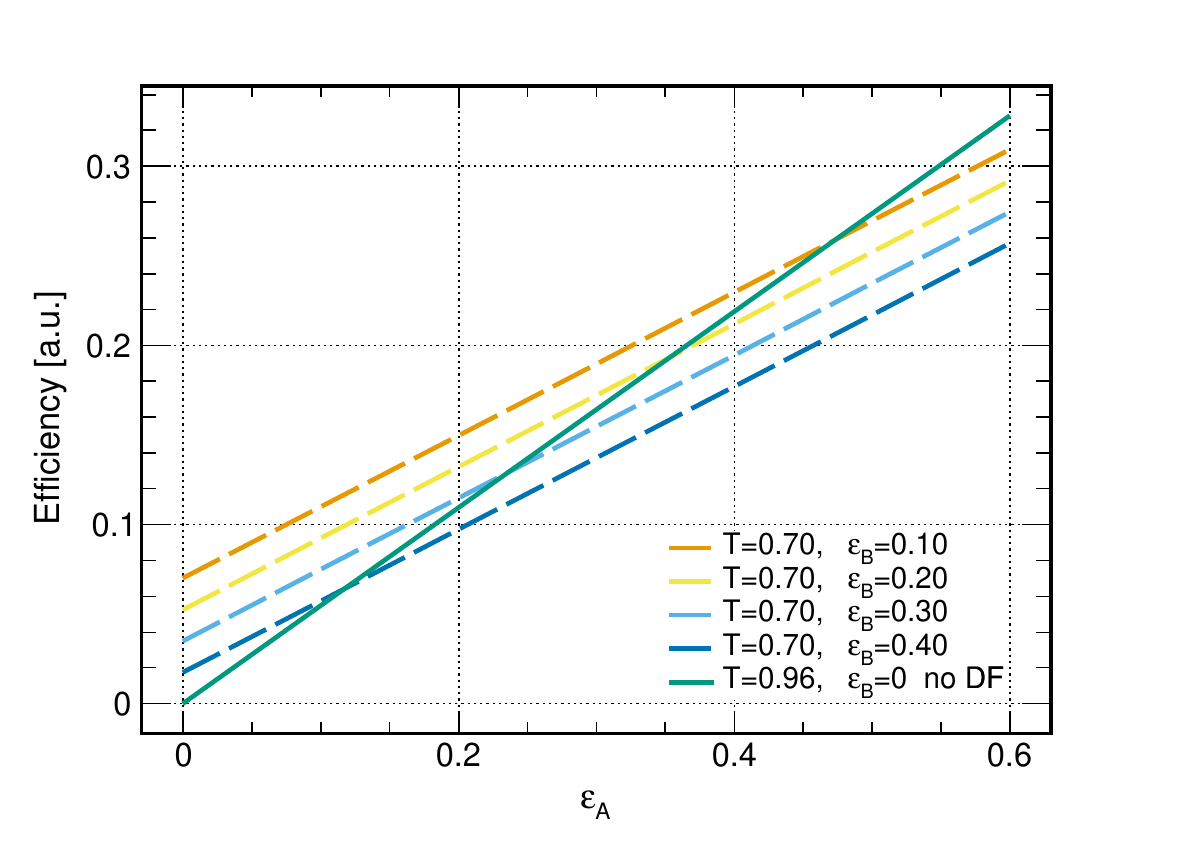}
    \caption{Overall Photon Collection Efficiency (PCE) as a function of PCE of the light guide $\epsilon_A$. The solid green line represents the module without dichroic filters, the dashed lines represent the module with dichroic filters with different collection efficiencies $\epsilon_B$.}
    \label{fig:PhColl}
\end{figure}
The effects of the optical window transparency and collection efficiencies $\epsilon_A$ and $\epsilon_B$ are shown in Figure~\ref{fig:PhColl}; the overall Photon Collection Efficiency (PCE) is plotted as a function of the light guide collection efficiency $\epsilon_A$ for different values of $T$ and $\epsilon_B$.
\begin{equation}
    PCE = (T \cdot 0.56)\epsilon_A + (T \cdot 0.25\cdot\epsilon_B)
\end{equation}
The dashed lines represent the PCE of the XA utilizing dichroic filters with arbitrary collection efficiencies $\epsilon_B$ between 10\% and 40\%, the presence of the filters also implies an optical window transparency $T=0.70$. The solid green line represent the PCE of the XA not utilizing dichroic filters, its $\epsilon_B$ is therefore zero but there's no penalty on the window transparency, that is $T=0.96$ in this case. It can be seen that, as $\epsilon_A$ grows, the XA configuration without dichroic filters gains advantage over the configuration with dichroic filters so, for a given dichroic collection efficiency $\epsilon_B$ and window transparency T, there is a light guide collection efficiency $\epsilon_A$ over which the adoption of dichroic filters is counterproductive.\\
The XA module with increased light sealing and two piece cut light guide, described in the previous section, allowed to collect more photons from the light guide, increasing its $\epsilon_A$ compared to the baseline configuration. We tested this configuration replacing the dichroic filters with clear glass substrates deposited with pTP: in Figure~\ref{fig:noDF_meas} the results are shown in comparison with the ones obtained previously.\\
The result obtained without utilizing the dichroic filters is compatible with the one employing them, this suggests the photons collected by the filter trapping mechanism compensate the loss caused by the decreased optical window transparency. In Figure~\ref{fig:PhColl} this situation would correspond to a point around the intersection of the lines representing the configurations with and without dichroic filters.\\
We repeated the same test with a baseline module replacing DF with clear substrates: in this case we obtained a lower PDE value ($\sim$-10\%), as it would be expected from the plot in Figure~\ref{fig:PhColl}, lowering $\epsilon_A$.\\
A measurement of the FD2-VD X-Arapuca PDE performed at CIEMAT \cite{MantheyCorchado_2025}, has shown that the presence of the DF-coated BF33 substrates caused the device to under perform of about 20\% with respect to the uncoated ones. This can be explained by the larger dimensions of the module (60~cm$~\times$~60~cm): the trapped photons have to travel more to reach the SiPMs, this severely penalizes the dicroic-coating transport (hence low $\epsilon_B$) while the WLS-LG transport is less affected ($\epsilon_B\ll\epsilon_A$); the effect of the lowered transparency to pTP photons is therefore enhanced in this configuration.
\section{Conclusions}
The tests performed at Milano-Bicocca on the FD-HD implementation of the X-Arapuca concept, give a better understanding of the X-Arapuca working principle. Dedicated measurements for the single components of the module allowed to disentangle the effect of each one on the overall performance. Thanks to this measurement campaign, with the help of the geant4 based Montecarlo simulation, it has been possible to optimize the FD-HD implementation of the X-Arapuca, increasing its photon detection efficiency up to a factor $\sim$84\%, from a PDE of $\sim$2.5\% to $\sim$4.6\%. The same information can be exploited to improve the performance of future implementations of the X-Arapuca such as the FD-VD one.\\
The optical coupling at the edges of the light guide proved to be a critical parameter, the overall PDE strongly depends on the gap between the SiPMs (and Vikuiti reflector) and the light guide (O(1\%) per O(0.1mm)).\\
The presence of the dichroic filter deposition on the inside of the entrance windows showed a $\sim$10\% improvement for the baseline configuration while no improvement has been observed in the improved configuration. This result clarifies the effect of the dichroic filter and gives valuable information on whether its employment is beneficial to the overall module performance.\\

\clearpage
\bibliography{bibliography.bib}

\end{document}